\documentstyle[seceq,preprint,eclepsf]{jpsj}

\newcommand{\CH}{{\cal H}}
\newcommand{\D}{{\cal D}}
\newcommand{\dtau}{{\rm d}\tau}
\newcommand{\drd}{{\rm d}^2r}
\newcommand{\drt}{{\rm d}^3r}
\newcommand{\dx}{{\rm d}x}

\newcommand{\dxt}{{\rm d}^3x}
\newcommand{\dxq}{{\rm d}^4x}

\newcommand{\dz} {{\rm d}z}
\newcommand{\dw} {{\rm d}\omega}
\newcommand{\ptau}{{\bf \tau}}
\newcommand{\pls}{\left(}
\newcommand{\prs}{\right)}
\newcommand{\pll}{\left[}
\newcommand{\prl}{\right]}
\newcommand{\pl}{\left\{}
\newcommand{\pr}{\right\}}
\newcommand{\al}{\left|}
\newcommand{\ar}{\right|}
\newcommand{\ple}{\left .}
\newcommand{\pre}{\right .}
\newcommand{\vn}{\mib{\nabla}}

\newcommand{\eff}{{\rm eff}}
\newcommand{\el}{{\rm el}}

\newcommand{\tr}{{\rm tr}}

\renewcommand{\Re}{{\rm Re}}
\renewcommand{\Im}{{\rm Im}}

\def\runtitle{Effect of Low Energy Excitations}
\def\runauthor{K. {\sc Awaka} and H. {\sc Fukuyama}}

\title
{Effects of Low Energy Excitations in SNS Junction on the 
Dynamics of Josephson Phase}

\author
{Kaoru {\sc Awaka}\footnote
{Present address: Texas Instruments Tsukuba Research 
and Development Center, Ltd, 17 Miyukigaoka, Tsukuba, 305}
\footnote{e-mail: awaka@trdc.ti.com}
and Hidetoshi {\sc Fukuyama}\footnote{e-mail: fukuyama@phys.s.u-tokyo.ac.jp}
}

\inst
{Department of Physics, Tokyo University, Tokyo 113}

\recdate
{
\today
}

\abst
{The effect of low energy excitation on the dynamics of 
Josephson phase in SNS junction is investigated. From the 
microscopic Hamiltonian, the effective action for the phase 
variable is derived. 
The retardation effects due to low energy excitation in normal(N) 
region are seen to play important roles in the dynamics of the phase. 
By the self consistent harmonic approximation, renormalization of mass 
and dissipation constant are calculated, revealing the enhancement of 
the former and the suppression of the latter in general. 
Various situation appears depending on these renormalized values.
}

\kword
{SNS junction, self-consistent harmonic approximation, 
retardation effect, tunneling}

\begin{document}
\sloppy
\maketitle
\section{Introduction}
The quantum mechanics of the phase of a superconductor in Josephson 
junction has been studied both theoretically and experimentally by many 
authors.~\cite{SZ1}
This problem was first treated phenomenologically in the pioneering work 
of Caldeira and Leggett based on the functional 
integral approach.~\cite{CL1,CL2}
Since the functional integral approach gives a clear view and the insight 
to the physics, it was developed thereafter to give the microscopic 
foundation to the quantum mechanics of the Josephson phase.~\cite{AES,EAS,LO1}
In these studies, the dynamics of the phase variable is described by the 
effective action, which can be derived from microscopic Hamiltonian by 
eliminating the electron field. 
It was shown that the effective action consists of the mass term 
which describes the charging energy, $E_c$, of the junction and the terms 
which express the effect of tunneling of quasi-particle and Cooper 
pair across the junction. 
The latter two terms, which express the effect of tunneling, 
can be viewed as the interaction between the phase at different imaginary 
times. 

In the case of superconductor-insulator-superconductor(SIS) junction with 
large capacitance, the mass of the phase is large and thus 
the time scale for the motion of the phase is long compared to that of 
the interaction term. Therefore the \lq\lq adiabatic" approximation can be 
justified and the kernel expressing the interaction between the phase at 
different imaginary time can be approximated by $\delta$-function.
With this approximation, the action reduces to that of a particle 
in a cosine potential with Ohmic dissipation. 
In this case, macroscopic quantum tunneling of the phase is predicted 
and experimentally observed.~\cite{CL1,LO2,GOW,VW,CMC}

For a SIS junction with small capacitance, 
Geigenm\"uller and Ueda~\cite{GU} investigated 
the non-adiabatic effect on the effective critical current of the junction by 
use of self consistent harmonic approximation(SCHA). 
They compared the results of SCHA with those of adiabatic approximation 
and find the qualitative agreement between the results obtained by 
the two approximations even if the charging energy is larger than the gap 
of the superconductor. 
Quantitatively, however, the difference between the adiabatic 
approximation and SCHA turns out to be quite large. 

As for the superconductor-normal metal-superconductor (SNS) junction, 
it is now possible to fabricate a junction with a capacitance small 
enough to observe macroscopic quantum effect.~\cite{THN1}
In contrast to SIS junction, low energy excitation in normal(N) region 
gives rise to the long time behavior of the kernel, $K(\tau)$, and hence 
$K(\tau)$ may not always be approximated by a $\delta$-function. 

The classical limit of the SNS junction with two dimensional electron gas(2DEG) 
with diffusive motion was investigated by Kresin,~\cite{Kresin}
who calculated the critical current, $j_c$.
In his approach, the charging energy, $E_c$, of the junction is assumed to be zero 
and, correspondingly, the phase valuable is independent of the imaginary time, 
{\it i.e.} $\theta(\tau)=\theta_0$.
In this case, Josephson coupling 
energy is given by $E_J=\int_0^\beta\dtau K(\tau)$, 
and the critical current, $j_c$, is obtained as $j_c=2eE_J$.
When the charging energy becomes finite, however, the quantum fluctuation of phase must 
be taken into account and we expect the retardation effect of $K(\tau)$ 
to appear in the dynamics of the phase.
In this paper, we will investigate this retardation effect 
on the tunneling rate of the phase in the neighborhood of critical current,
where the tunneling barrier height becomes small. 

Apart from the problem of Josephson junction, the tunneling rate in the 
small barrier height region was studied by Kleinert when there is 
no dissipation and the retardation effect.~\cite{Kleinert1}
In the present  paper, we will extend his method 
to the case with the dissipation and retardation effect 
to calculate the tunneling rate in the neighborhood of the 
critical current. 

To investigate how the retardation of the kernel affects 
the decay rate of the metastable state,
we first derive the effective action for the Josephson phase in SNS junction 
from the microscopic Hamiltonian, following the procedure outlined by 
Ambegaokar {\it et al.}~\cite{AES,EAS}.
Then we apply self consistent harmonic approximation
 to the action, and calculate the renormalization of 
the mass, dissipation constant and attempt frequency.
With these renormalized parameters, we will study the tunneling rate in the
neighborhood of the critical current.

The organization of the paper is as follows.
In the next section, we will derive the effective action for Josephson phase 
in SNS junction in the perturbation theory with respect to the mixing matrix 
element between S and N region.
In section 3, we will explain the self-consistent harmonic approximation 
applied to the effective action derived in section 2.
The results and discussion will be given in section 4 and 
a brief summary in section 5.

We take units, $\hbar=k_B=1$, unless noted.

\section{Effective Action for SNS Junction}
In this section, we will derive the effective action from microscopic 
Hamiltonian following Ambegaokar {\it et al.}\cite{AES}

\subsection{Derivation of the Effective Action}
\begin{figure}
\epsfile{file=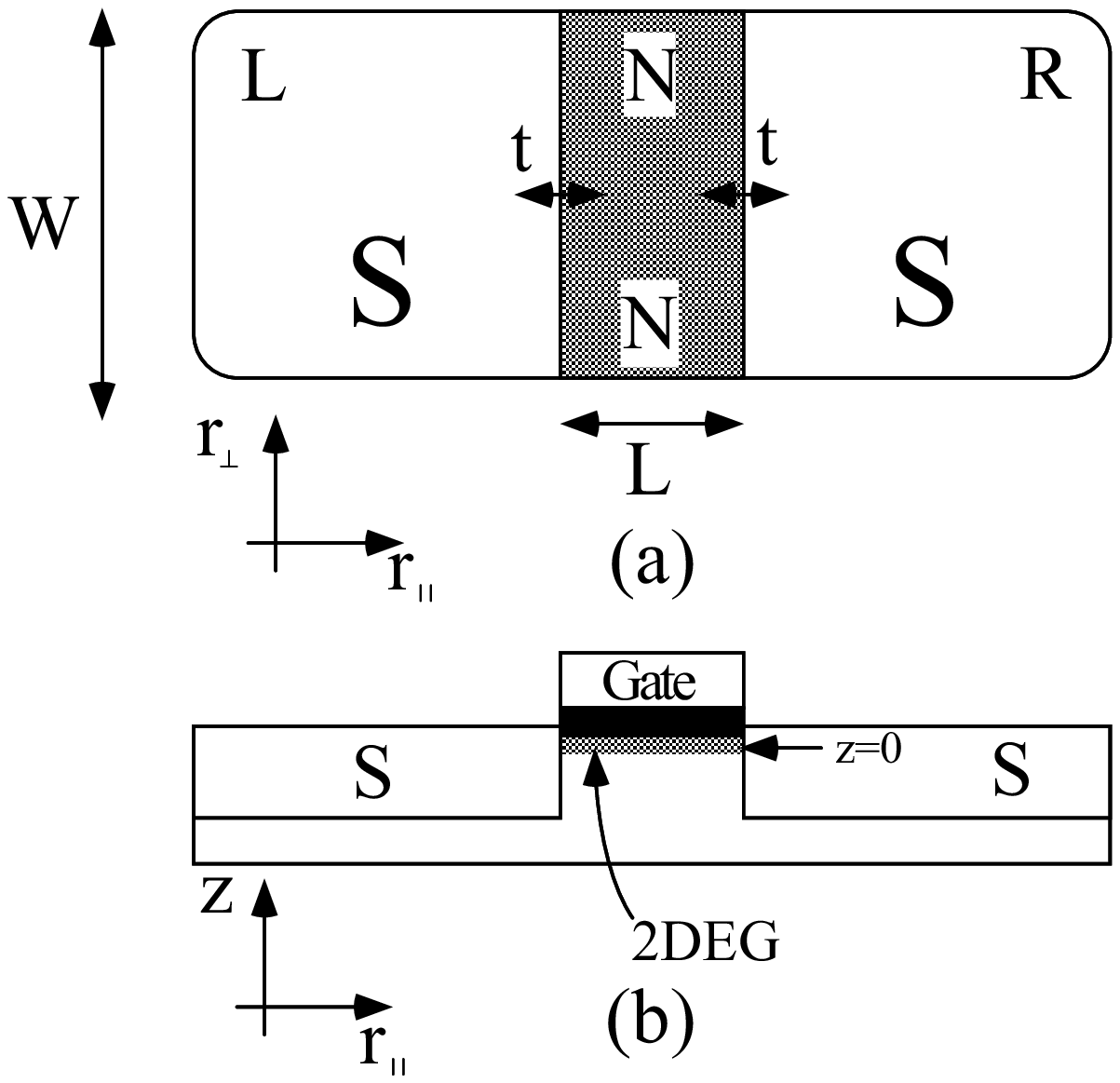,width=7cm}
\caption{A schematic picture of SNS junction. The top view(a) and the 
cross sectional view(b) are shown. 
The superconducting(S) regions, L and R, are connected to 2DEG of
normal(N) region by the transfer integral $t_{rr'}$. The width and 
the length of the junction are $W$ and $L$, respectively.
The 2DEG plane lies at $z=0$.}
\label{model}
\end{figure}

The model of the junction is as follows(See Fig.~\ref{model}):
Two superconducting regions(S region), which will be referred as $L$-
and $R$-region, are connected by a normal region(N region) of length, $L$, 
and width, $W$.
The S regions are assumed to be clean bulk superconductor and
N region to be 2DEG with impurities.
Finite transfer integral, $t_{rr'}$,  of the electrons is
assumed through the boundaries of S and N regions.
The transfer integral can be taken as real without a loss of generality.
We will take $z$ axes perpendicular to the 2DEG. Other axes, $r_\parallel$ 
and $r_\perp$ are taken to be parallel to and perpendicular to the junction 
direction, respectively, as shown in Fig.~\ref{model}.

%#### Figure ####
%################

The Hamiltonian of the system is given as follows.
\begin{full}
\begin{subeqnarray}
\CH&=&\CH_{\rm L}+\CH_{\rm N}+
\CH_{\rm R}+\CH_{\rm T}+\CH_{\rm em},\\
\CH_{\rm L}&=&\int_{r\in{\rm L}}\drt\,\psi^*_{{\rm L}\sigma}\pls r\prs
\pll-{1\over2m_e}{\mib\nabla}^2-\mu-e\phi\pls r\prs\prl
\psi^{\phantom *}_{{\rm L}\sigma}\pls r\prs-{g\over2}\int_{r\in L}\drt
\psi^*_{{\rm L}\uparrow}\pls r\prs
\psi^*_{{\rm L}\downarrow}\pls r\prs
\psi^{\phantom *}_{{\rm L}\downarrow}\pls r\prs
\psi^{\phantom *}_{{\rm L}\uparrow}\pls r\prs,\ \ \ \ \ \ \ \\
\CH_{\rm R}&=&({\rm L}\leftrightarrow{\rm R}),\\
\CH_{\rm N}&=&\int_{r\in {\rm N}}{\rm d}^2r\,\psi^*_{{\rm N}\sigma}
\pls r\prs\pll-{1\over2m_e}{\mib \nabla}^2
-\mu-e\phi(r)-U(r)\prl\psi^{\phantom *}_{{\rm N}\sigma}
\pls r\prs,\\
\CH_{\rm T}&=&
-\left(\int_{r\in {\rm L}, r'\in {\rm N}}\drt{\rm d}^2r'\,t_{r,r'}
\psi^*_{{\rm L}\sigma}\pls r\prs\psi^{\phantom{*}}_{{\rm N}\sigma}\pls r'\prs
+{\it h.c.}\right)\nonumber\\
&\phantom{=}&-\left(\int_{r\in {\rm R}, r'\in {\rm N}}\drt{\rm d}^2r'\
\,t_{r, r'}\psi^*_{{\rm R}\sigma}\pls r\prs\psi_{{\rm N}\sigma}\pls r'\prs
+{\it h.c.}\right),\\
\CH_{\rm em}&=&{1\over8\pi}\int\drt
\pll{\mib \nabla}\phi(r)\prl^2,
\label{2.1}
\end{subeqnarray}
\end{full}
where $g$, $m_e$, $\mu$ and $-e$ are  interaction constant,
effective mass of an electron, chemical potential and electron
charge, respectively.
The electron field and scalar potential are expressed by $\psi(r)$ and 
$\phi(r)$, respectively.
The impurities in N region is expressed by a potential $U(r)$, 
which is assumed to be randomly distributed delta function,
$U(r)=\sum_j^{N_i} V_{\rm imp}\delta(r-r_j)$.
Here $N_i$, $V_{\rm imp}$ and $r_j$ are the number of impurities, 
the strength of the impurity potential and the position of the j-th 
impurity, respectively.
Since the wave function of the electrons in N-region is restricted to a 
narrow region around $z=0$ as in Fig.~\ref{model}b, 
the scalar potential in N-region is evaluated at $z=0$. 
In eq.(\ref{2.1}), the effect of vector potential is not explicitly taken 
into account, because it is not necessary in the following discussions.

The action of the system is defined by,
\begin{eqnarray}
Z&\equiv&\tr\ e^{-\beta\CH}\nonumber\\
&=&\int\D \psi^*\D\psi^{\phantom *}\D\phi\ e^{-S}.
\label{2.2}
\end{eqnarray}
In eq.(\ref{2.2}), the Wick rotation is performed, {\it i.e.} 
${\rm i}t\to\tau$ and $\phi\to {\rm i}\phi$.

First, we eliminate the quartic interaction term in Hamiltonian, eq.(\ref{2.1}), by the Hubbard-Stratonovich transformation,
\begin{full}
\begin{equation}
1=\int\D\Delta^*\D\Delta\,
\exp\pll-{1\over2g}\int\dx
\pls\Delta(x)-g\psi_{\downarrow}(x)\psi_{\uparrow}(x)\prs
\pls\Delta^*(x)-g\psi^*_{\uparrow}(x)
\psi^*_{\downarrow}(x)\prs
\prl,
\end{equation}
\end{full}
where $x$ denotes both imaginary time, $\tau$, and
position, $r$.
With this transformation, the action becomes,
\begin{full}
\begin{subeqnarray}
S&=&S_{\rm cond}+S_{\rm em}+S_{\rm el},\\
S_{\rm cond}&\equiv&\int{\rm d}x{|\Delta_{\rm L}|^2\over2g}
+\int{\rm d}x{|\Delta_{R}|^2\over2g},\\
S_{\rm em}&\equiv&{1\over8\pi}\int{\rm d}x
\pll{\mib \nabla} \phi\pls x\prs\prl^2,\\
S_{\rm el}&\equiv&\int{\rm d}x{\rm d}x'\Psi^{*}\pls x\prs
\pll-G^{-1}\pls x, x'\prs\prl\Psi^{\phantom *}\pls
x'\prs.
\label{2.4}
\end{subeqnarray}
\end{full}
The Green's function, $G\pls x, x'\prs$, is a $6\times6$
matrix spanned both in the space of L, N, R and Nambu space.
The electron fields are expressed as,
\begin{subeqnarray}
\Psi\pls x\prs &\equiv&\pll
\matrix{
\Psi_{\rm L}\pls x\prs\cr
\Psi_{\rm N}\pls x\prs\cr
\Psi_{\rm R}\pls x\prs\cr}\prl,\\
\Psi_{{\rm L(N, R)}}\pls x\prs&=&\pll
\matrix{
\psi_{{\rm L(N, R)}\uparrow}\pls x\prs\cr
\psi_{{\rm L(N, R)}\downarrow}\pls x\prs\cr}\prl.
\end{subeqnarray}
The inverse of the Green's function for electron system,
$G^{-1}\pls x, x'\prs$, is given by,
\begin{full}
\begin{equation}
G^{-1}\pls x, x'\prs
=\pll
\matrix{
G^{-1}_L\pls x, x'\prs&-t_{xx'}\tau_3&0\cr
-t_{xx'}\tau_3&G^{-1}_N\pls x, x'\prs &-t_{xx'}
\tau_3\cr
0&-t_{xx'}\tau_3&G^{-1}_R\pls x, x' \prs\cr
}\prl.
\end{equation}
\end{full}
Here the diagonal elements are,
\begin{full}
\begin{subeqnarray}
G_{{\rm L(R)}}^{-1}\pls x, x'\prs
&=&\pll{-\partial_\tau}{\bf 1}
-\pls -{1\over2m_e}{\mib \nabla}^2
-\mu-{\rm i} e \phi \prs \ptau_3+
\Delta e^{{\rm i} \theta_{\rm L(R)}\ptau_3}\ptau_1 \prl
\delta\pls x-x'\prs,\\
G_{\rm N}^{-1}\pls x, x'\prs
&=&\pll {-\partial_\tau{\bf 1}}
-\pls -{1\over2m_e}{\mib \nabla}^2-\mu-{\rm i} e \phi+U\prs
\ptau_3 \prl\delta\pls x-x'\prs,
\label{2.7}
\end{subeqnarray}
\end{full}
where $\ptau_1$, $\ptau_3$ are the Pauli matrices and $\bf 1$ is a unit
matrix and $\theta_{L(R)}$ is the phase of the superconducting order 
parameter, $\Delta_{L(R)}=\Delta {\rm e}^{i\theta_{L(R)}}$.

Before tracing out the electron degrees of freedom from the
action, eq.(\ref{2.4}), we perform a transformation,
\begin{equation}
\Psi_{L(N, R)}\pls x\prs=
{\rm e}^{{\rm i}{1\over2}\theta_{L(N, R)}
\pls x\prs \tau_3}
{\tilde \Psi}_{L(N, R)}\pls x\prs,
\end{equation}
to express the phase explicitly and then eliminate them from the order 
parameters in eq.(\ref{2.7}).
Here we have also introduced a phase in normal region which 
satisfies~\cite{Sch1},
\begin{equation}
{\partial\over\partial\tau}\theta_{\rm N}(x,\tau)
=2e\phi\pls x, \tau\prs,
\label{2.9}
\end{equation}
to get rid of the scalar potential in N-region. 
Note that the scalar potential in eq.(\ref{2.9}) is 
considered as $\phi(x)=\phi(r_\parallel,r_\perp,z=0,\tau)$, since 
the electron gas in N-region is confined in the narrow region around $z=0$. 
With eq.(\ref{2.9}) the 
phase can be defined continuously from S-region to N-region at the boundary.
(See eq.(\ref{2.16}.))
Then we arrive at the action,
\begin{equation}
S_{\rm el}=\int \dx\dx'{\tilde \Psi}^*\pls x\prs
\pll-{\tilde G}^{-1}\pls x, x'\prs \prl
{\tilde \Psi}^{\phantom *}\pls x'\prs,
\label{2.10}
\end{equation}
where the inverse of Green's function is expressed as,
\begin{full}
\begin{equation}
{\tilde G}^{-1}\pls x, x'\prs
=\pll
\matrix{
{\tilde G}^{-1}_L\pls x,x'\prs&-t_{xx'}\tau_3&0\cr
-t_{xx'}\tau_3&{\tilde G}^{-1}_N\pls x,x'\prs &
-t_{xx'}\tau_3\cr
0&-t_{xx'}\tau_3&{\tilde G}^{-1}_R\pls x, x'\prs
}
\prl.
\end{equation}
\end{full}
The diagonal elements are,
\begin{full}
\begin{equation}
{\tilde G_{L(N,R)}(x,x')\equiv
{\rm e}^{-{\rm i}{1\over2}\theta_{L(N, R)}(x)\tau_3}}
G_{L(N,R)}(x,x')
{\rm e}^{{\rm i}{1\over2}\theta_{L(N, R)}(x)\tau_3}.
\label{2.12}
\end{equation}
\end{full}
Tracing out the electron fields from eq.(\ref{2.10}),
we obtain the action as follows to the second order of the matrix element of 
the transfer integral,
\begin{full}
\begin{eqnarray}
S_{\rm el}&=&
-\tr\log\pll -{\tilde G}^{ -1}\pls x,x'\prs\prl
\nonumber\\
&=&-\tr\log\pll -{\tilde G}_0^{-1}\prl
-{1\over2}\tr\log\pll{\bf 1}
-{\tilde G}_0{\bf T}{\tilde G}_0{\bf T}\prl\nonumber\\
&\sim&
-\tr\log\pll-{\tilde G}_0^{-1}\prl
+{1\over2}\tr\pll {\tilde G}_0{\bf T}{\tilde G}_0{\bf T}
+{1\over2}{\tilde G}_0{\bf T}{\tilde G}_0{\bf T}
{\tilde G}_0{\bf T}{\tilde G}_0{\bf T}\prl.
\label{2.13}
\end{eqnarray}
\end{full}
Here 
\begin{full}
\begin{subeqnarray}
{\tilde G}_0\pls x, x'\prs&=&\pll
\matrix{
{\tilde G}_L\pls x, x'\prs&0&0\cr
0&{\tilde G}_N\pls x, x'\prs&0\cr
0&0&{\tilde G}_R\pls x, x'\prs}\prl,
\\
{\bf T}&=&\pll
\matrix{
0&-t_{xx'}\tau_3&0\cr
-t_{xx'}\tau_3&0&-t_{xx'}
\tau_3\cr
0&-t_{xx'}\tau_3&0\cr}\prl.
\end{subeqnarray}
\end{full}
In eq.(\ref{2.13}), $\tr$ means taking trace with respect to the matrix
element of the Green's function and integrating with respect to $x$,$x'$.

Now the action, $S$, depends on $\Delta$, $\phi$ 
and $\theta$, which are independent fluctuating variables.
Fortunately, these quantities, except the phase $\theta$,
have their mean field values with only small
fluctuations around them, which we ignore.

First we investigate the zero-th order term of eq.(\ref{2.13}).
The action for $L$, $N$, $R$ region are defined by,
\begin{equation}
S_{L(N,R)}\equiv-\tr\log\pll -{\tilde G}_{L(N, R)}
\pls x, x'\prs\prl.
\end{equation}
Expanding the Green's function in L- and
R-region with respect to $\vn\theta(x)$
and $\partial_\tau \theta(x)-{\rm i} e \phi(x)$ to the second
order, we obtain as the relevant part of the effective
action for S-region,
\begin{full}
\begin{eqnarray}
S_{L(R)}&\sim&
{N_0\over\Omega_{L(R)}}
\int_{x\in L(R)}
\dxq\pll {1\over2}
{\partial\theta\pls x\prs\over\partial\tau}-e \phi\pls x\prs\prl^2
+\int_{x\in L(R)} \dxq
{m \rho_s\over 2}\pll{1\over2m}
{\mib \nabla} \theta\pls x\prs\prl^2,
\label{2.15}
\end{eqnarray}
\end{full}
where $\Omega_{\rm L(R)}$ is the volume of the superconductor, $\rho_s$
is the superfluid density and $N_0$ is the density of states per spin 
at Fermi energy.~\cite{AES}

From eq.(\ref{2.15}),
we see that the action is extremal when
the phase variable satisfies the Josephson relation,~\cite{Ti:text}
\begin{equation}
{\partial\over\partial \tau}
\theta\pls x\prs 
=2e\phi\pls x \prs.
\label{2.16}
\end{equation}
With eqs.(\ref{2.9}) and (\ref{2.16}), the phase of the S region and that of
the N region are connected smoothly at the boundary.

On the other hand in N-region, by expanding the action with
$\vn \theta(x)$, we obtain,
\begin{full}
\begin{equation}
S_N\sim
{1\over2}\int_{x,x'\in N}\dxt\dxt'{1\over2}\vn_\alpha
\theta\pls x \prs \pi^{(\perp)}_{N,\alpha\beta}
\pls x,x'\prs
{1\over2}\vn_\beta \theta\pls x'\prs
+\int_{x\in N}\dxt {\rho_0\over2m}\pll {1\over2}\vn\theta
\pls x\prs\prl^2,
\label{2.16.5}
\end{equation}
\end{full}
where 
$\pi^{(\perp)}_{N,\alpha\beta}\pls x,x'\prs\equiv
-\langle T_{\tau}j_\alpha(r,\tau)j_\beta(r',\tau')\rangle$ 
is current correlation function in N region, $\rho_0$ is the density
of the electrons and indices $\alpha,\beta$ stand for $r_\perp$ and
$r_\parallel$. Here $T_{\tau}$ is the time ordering operator and $j_\alpha(r,\tau)$ is 
the current operator. Since we have taken the phase which satisfies eq.(\ref{2.9}), 
the contribution from density correlation function does not appear 
in eq.(\ref{2.16.5}).
Noting that the characteristic decay length of 
$\pi^{(\perp)}_{N,\alpha\beta}\pls x,x'\prs$ is the electron mean free path, 
$l$, and the variation of the scalar potential is small, $\theta(x)$ and 
$\theta(x')$ in the first term of eq.(\ref{2.16.5}) can be evaluated 
at the same point, {\it i.e.}, $r=r'$. 
By performing integration with respect to $r'$ in 
$\pi^{(\perp)}_{N,\alpha\beta}(x,x')$,
the effective action in eq.(\ref{2.16.5}) can be approximated as,
\begin{full}
\begin{eqnarray}
S_N&\sim&
{1\over2}\int_{x, x'\in N}\dtau\dtau'{\rm d}^2r{\rm d}^2r'{1\over2}\vn_\alpha
\theta\pls r, \tau \prs \pi^{(\perp)}_{N,\alpha\beta}
\pls x,x'\prs
{1\over2}\vn_\beta \theta\pls r,\tau'\prs
+\int_{x\in N}\dtau{\rm d}r {\rho_0\over2m}\pll {1\over2}\vn\theta
\pls r,\tau \prs\prl^2,\nonumber\\
&\equiv&\int\dtau\dtau'\drd
\pll{1\over2}\vn_\alpha\theta(r,\tau)\prl
\pi (\tau, \tau')
\pll{1\over2}\vn_\alpha\theta(r,\tau')\prl,
\label{2.20}
\end{eqnarray}
\end{full}
where the correlation function, $\pi(\tau,\tau')$, is given by,
\begin{eqnarray}
\pi\pls \tau, \tau' \prs
&=&{1\over 2}\int{\rm d}^2r' \pi^{(\perp)}_{N,\alpha\alpha}(r,r';\tau,\tau')
+{\rho_0\over 2m}\delta(\tau-\tau')\nonumber\\
&=&{\rho_0\tau_\tr\over2m}
{1\over\beta}\sum_n\,{\rm e}^{-i\omega_n(\tau-\tau')}{\al\omega_n\ar},
\end{eqnarray}
where $\tau_\tr$ is the transport scattering time of the electron
due to elastic scattering and $\omega\tau_{\tr}\ll 1$ is assumed. 
Weak localization effect is not taken into account here.

In terms of the phase difference,
\begin{equation}
\theta^-(\tau)=\theta(x_L, \tau)-\theta(x_R, \tau),
\end{equation}
which is assumed to be uniform in $r_\perp$
direction, eq.(\ref{2.20}) becomes,
\begin{equation}
S_N\sim{1\over2\beta}
\sum_n{\alpha\over2\pi}|\omega_n|\theta^-_n\theta^-_{-n},
\label{2.21}
\end{equation}
where
$\alpha=R_Q/R_N$. Here $R_Q={h/(4e^2)}\sim6.45 {\rm k\Omega}$
is the quantum resistance and
$R_N=L/(W\sigma)$ is the resistance of the junction.
In eq.(\ref{2.21}), we assumed 
${\mib \nabla}\theta\pls r,\tau\prs\sim\theta_-\pls \tau\prs/L$,
which means the electric field is constant in N region.
The term eq.(\ref{2.21}) has been derived by Sch\"on and Zaikin 
for SNS junction.\cite{SZ2}

\begin{figure}
\epsfile{file=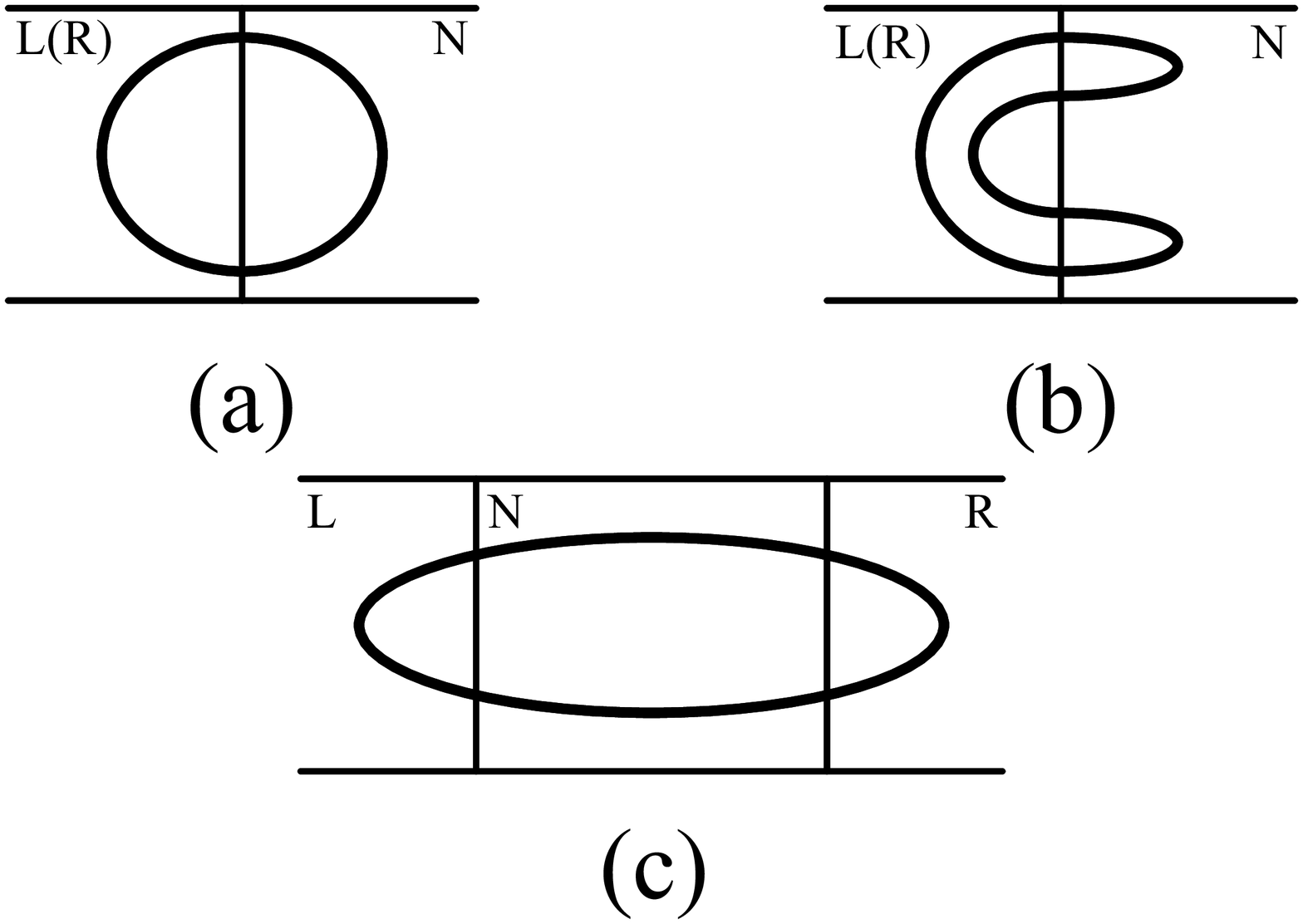,width=7cm}
\caption{The diagrams which appears in the expansion of eq.(2.13)
with respect to the matrix element of the transfer integral $t$.}
\label{diagrams}
\end{figure}

Next we investigate the effect of the matrix element of the transfer 
integral in eq.(\ref{2.13}).
The diagrams expressing these processes are listed in Fig.~\ref{diagrams}

%#### Figure ####
%################

The lowest order contribution(See Fig.~\ref{diagrams}a), is given as,
\begin{full}
\begin{eqnarray}
S_2&\equiv&
{1\over2}\tr\pll{\tilde G}_0{\bf T}{\tilde G}_0
{\bf T}\prl\nonumber\\
&=&{1\over2}\tr\pll{\tilde G}_L\pls
x, x'\prs t_{x'x}\ptau_3{\tilde G}_N\pls x', x
\prs t_{xx'}\ptau_3+{\tilde G}_R\pls x, x'
\prs t_{x'x}\ptau_3{\tilde G}_N\pls x', x\prs
t_{xx'}\ptau_3\prl\nonumber\\
&=&t^2\int\dx\dx'{\tilde g}_{L\uparrow}\pls x, x'
\prs {\tilde g}_{N\uparrow}\pls x', x\prs
+{\tilde g}_{L\downarrow}\pls x, x'\prs
{\tilde g}_{N\downarrow}\pls x', x\prs
+\pll L\to R\prl.\label{and}
\end{eqnarray}
\end{full}
Here we made use of the fact that the matrix element of the 
transfer integral,
$t_{xx'}$, differs from $0$ only for $x\sim x'$ and in
the neighborhood of the S-N boundary, {\it i.e.}
$t_{xx'}=t\delta(r_\parallel-{r'}_\parallel)\delta(r_\perp-{r'}_\perp)
\delta(r_\parallel-r_\parallel^b)\delta(z)$. Here $r_\parallel^b$ is the 
position of boundary between R(L) and N-region. 
Therefore the integral in
eq.(\ref{and}) is taken over $\tau$ and $r_\perp$.
The Green's function ${\tilde g}_{\uparrow}$ and
${\tilde g}_{\downarrow}$ in eq.(\ref{and}) are the 1,1- and 2,2-
component of Nambu Green's function, respectively.
Using eqs.(\ref{2.7}a), (\ref{2.12}) and (\ref{2.16}), we find,
\begin{full}
\begin{equation}
{\tilde G}_{\rm L(R)}^{-1}\pls x, x'\prs=
\pll{-\partial_\tau}{\bf 1}-\pls -{1\over2m}\vn^2
-\mu\prs \ptau_3+\Delta\ptau_1 \prl
\delta\pls x-x'\prs.
\end{equation}
\end{full}
Due to the existence of the gap,
$\Delta$, in the excitation
spectrum, the Green's function becomes short ranged in time and space.
Thus we approximate it as, $G(x-x')\propto\delta\pls x-x'\prs$.

\begin{figure}
\epsfile{file=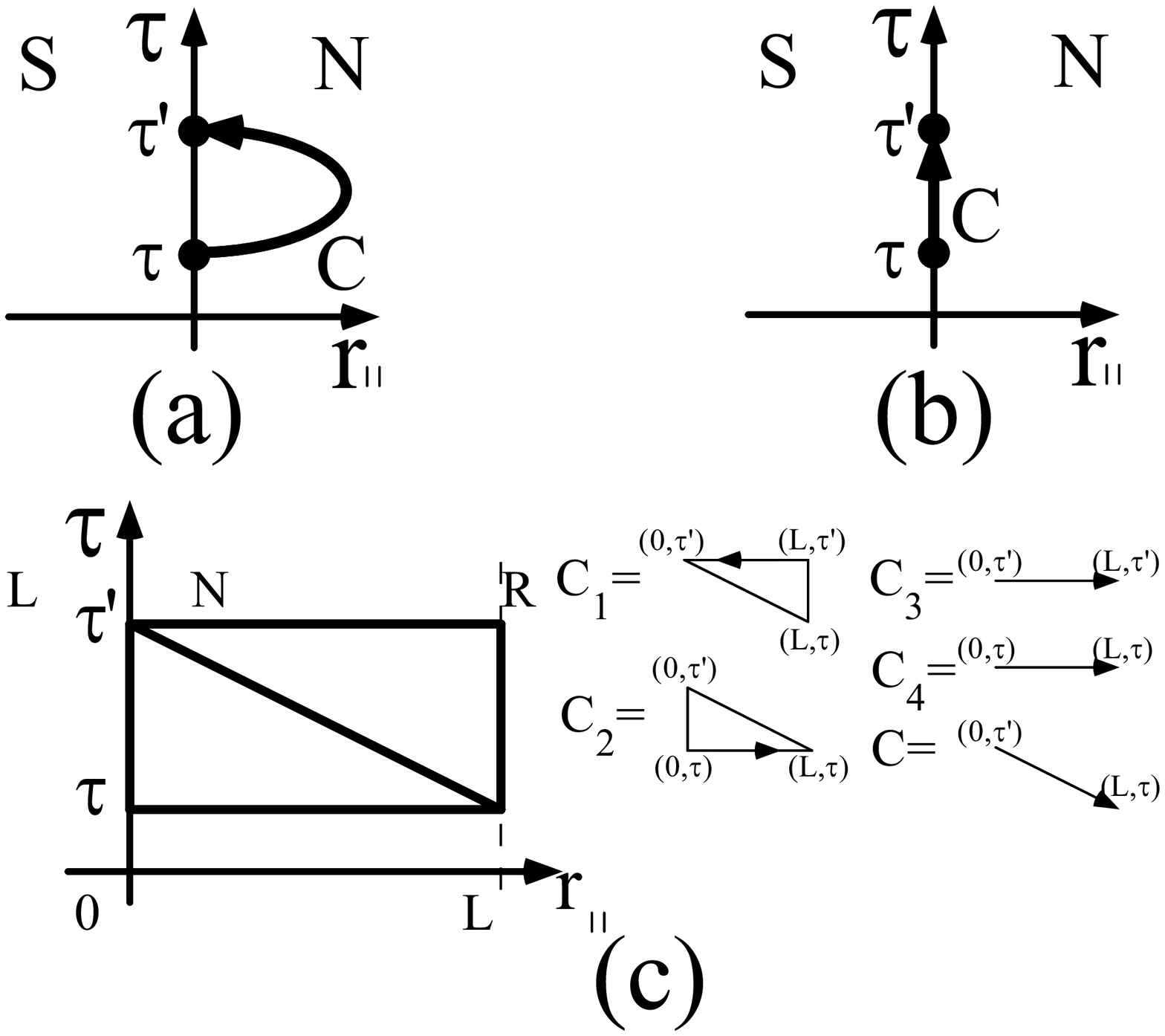,width=7cm}
\caption{Contours for the integrals in (a) eq.(2.25),
(b) eq.(2.27) and (c) eq.(2.30b).}
\label{contour}
\end{figure}

We also use the quasi-classical approximation to the
Green's function in N region, which leads to,
\begin{full}
\begin{equation}
{\tilde G}_{\rm N}\pls x, x'\prs\sim\exp\pll i\int_C {\rm d}
{\vec x}\cdot\pls {1\over2}{\partial\theta\over
\partial\tau}-e\phi, {1\over2}\vn
\theta\prs \ptau_3\prl
G_{\rm N}\pls x, x'\prs,
\label{2.25}
\end{equation}
\end{full}
where 
${\rm d}{\mib x}\cdot(a,{\mib v})\equiv a{\rm d}\tau
+{\mib v}\cdot{\rm d}{\mib r}$ and the path C is shown in Fig.~\ref{contour}a.
Since $G_L\pls x, x'\prs\propto\delta\pls x-x'\prs$,
the phase contribution in eq.(\ref{2.25}) disappears.
Thus the term shown in Fig.~\ref{diagrams}a is independent of 
the phase.

%#### Figure ####
%################

Next we turn to the fourth order diagrams,
{\it i.e.} Fig.~\ref{diagrams}b and Fig.~\ref{diagrams}c.
The first diagram is expressed as,
\begin{full}
\begin{eqnarray}
S_A&\equiv&{t^4\over2}\int\dx_1\dx_2\dx_3\dx_4
{\tilde g}_{L\uparrow}\pls x_1, x_2 \prs
{\tilde g}_{N\uparrow}\pls x_2, x_3 \prs
{\tilde g}_{L\uparrow}\pls x_3, x_4 \prs
{\tilde g}_{N\uparrow}\pls x_4, x_1\prs\nonumber\\
&\phantom{=}&+
{\tilde g}_{L\downarrow}\pls x_1, x_2 \prs
{\tilde g}_{N\downarrow}\pls x_2, x_3 \prs
{\tilde g}_{L\downarrow}\pls x_3, x_4 \prs
{\tilde g}_{N\downarrow}\pls x_4, x_1\prs\nonumber\\
&\phantom{=}&+
{\tilde f}_{L}\pls x_1, x_2 \prs
{\tilde g}_{N\downarrow}\pls x_2, x_3 \prs
{\tilde f}_{L}^*\pls x_3, x_4 \prs
{\tilde g}_{N\uparrow}\pls x_4, x_1\prs\nonumber\\
&\phantom{=}&+
{\tilde f}_{L}^*\pls x_1, x_2 \prs
{\tilde g}_{N\downarrow}\pls x_2, x_3 \prs
{\tilde f}_{L}\pls x_3, x_4 \prs
{\tilde g}_{N\uparrow}\pls x_4, x_1 \prs.
\end{eqnarray}
Here $f, f^*$ denote 1,2- and 2,1- component of the Green's function.
The first and the second term do not contain the
phase variable,
because the contributions from $g_{N\uparrow}(x, x')$
and $g_{N\uparrow}(x', x)$ cancels each other.
In the third and the forth term, the phase factor is
given by,
\begin{equation}
2i\int_C{\rm d}{\mib x}\cdot
\pls{1\over2}{\partial\theta\over\partial\tau}-e\phi,
{1\over2}{\mib \nabla}\theta\prs,
\label{2.27}
\end{equation}
\end{full}
with the path shown in Fig.~\ref{contour}b.
In principle, the contour, $C$, for the integrand must be
determined from the extremal path for the electron.
Here we approximate it with the line connecting the terminal points, $x, x'$ 
as shown in Fig.\ref{contour}b.
Noting the Josephson relation, eq.(\ref{2.9}), and that the
junction is uniform in $r_\perp$ direction, the phase contribution in
eq.(\ref{2.27}) will vanish.
This reflects the assumption we have made that there is no voltage drop
across the S-N boundary.
So the diagram in Fig.~\ref{diagrams}b can also be neglected.

The last diagram, Fig.~\ref{diagrams}c, is the only relevant term in our
analysis.
From this, we obtain the action as
\begin{full}
\begin{eqnarray}
S_J&\equiv&
{t^4\over2}\int\dx_1\dx_2\dx_3\dx_4
{\tilde g}_{L\uparrow}\pls x_1, x_2 \prs
{\tilde g}_{N\uparrow}\pls x_2, x_3 \prs
{\tilde g}_{R\uparrow}\pls x_3, x_4 \prs
{\tilde g}_{N\uparrow}\pls x_4, x_1\prs\nonumber\\
&\phantom{=}&+
{\tilde g}_{L\downarrow}\pls x_1, x_2 \prs
{\tilde g}_{N\downarrow}\pls x_2, x_3 \prs
{\tilde g}_{R\downarrow}\pls x_3, x_4 \prs
{\tilde g}_{N\downarrow}\pls x_4, x_1\prs\nonumber\\
&\phantom{=}&+
{\tilde f}_{L}\pls x_1, x_2 \prs
{\tilde g}_{N\downarrow}\pls x_2, x_3 \prs
{\tilde f}_{R}^*\pls x_3, x_4 \prs
{\tilde g}_{N\uparrow}\pls x_4, x_1\prs\nonumber\\
&\phantom{=}&+
{\tilde f}_{L}^*\pls x_1, x_2 \prs
{\tilde g}_{N\downarrow}\pls x_2, x_3 \prs
{\tilde f}_{R}\pls x_3, x_4 \prs
{\tilde g}_{N\uparrow}\pls x_4, x_1 \prs\nonumber\\
&\phantom{=}&+
\pll L\leftrightarrow R\prl.
\label{2.28}
\end{eqnarray}
\end{full}
In eq.(\ref{2.28}), we find the third and the forth terms, which
express Cooper pair transfer through the N region, are relevant and lead to,
\begin{full}
\begin{eqnarray}
S_J&=&t^4\int\dx_1\dx_2\dx_3\dx_4\pll
{\tilde f}_{L}^*\pls x_1, x_2 \prs
{\tilde g}_{N\downarrow}\pls x_2, x_3 \prs
{\tilde f}_{R}\pls x_3, x_4 \prs
{\tilde g}_{N\uparrow}\pls x_4, x_1 \prs\pre\nonumber\\
&\phantom{=}&+
\ple{\tilde f}_{R}^*\pls x_1, x_2 \prs
{\tilde g}_{N\downarrow}\pls x_2, x_3 \prs
{\tilde f}_{L}\pls x_3, x_4 \prs
{\tilde g}_{N\uparrow}\pls x_4, x_1 \prs\prl.
\end{eqnarray}
\end{full}
For low energy, we obtain,
\begin{full}
\begin{subeqnarray}
S_J&=&
-\int{\rm d}r_{\perp}{\rm d}r'_{\perp}{\rm d}\tau{\rm d}\tau'
K(x,x')\cos\pll W_C\pls x, x'\prs\prl,\\
W_C&\equiv&2\int_C {\rm d}{\vec x}\cdot
\pls {1\over2}{\partial\theta\over\partial\tau}-e\phi, 
{1\over2}\vn \theta\prs,\\
K(x, x')&\equiv&-2t^4\pll\int{\rm d}r_{\perp}{\rm d}\tau 
f_R(x)\prl
\pll \int{\rm d}r_{\perp}{\rm d}\tau f_L^*(x)\prl
\langle g_{N\uparrow}(x, x')
g_{N\downarrow}(x',x)\rangle_{\rm imp},
\label{2.29}
\end{subeqnarray}
\end{full}
where the contour, C, is taken as in Fig.~\ref{contour}c and
$\langle\cdots\rangle_{\rm imp}$ denotes impurity average.
The integral region in eqs.(\ref{2.29}a) and 
(\ref{2.29}c) are on the SN boundary.
Now we consider the case when the time dependence of the electric
field is weak.
In this case, the contour can be approximated by the average of 
C3 and C4 of Fig.~\ref{contour}c,
since the contribution from path
C1 and C2 in Fig.~\ref{contour}c cancels each other.
Thus we obtain,
\begin{equation}
W_C={\theta^-(\tau)+\theta^-(\tau')\over2}.
\end{equation}
With this approximation, the action becomes,
\begin{full}
\begin{equation}
S_J=-\int_0^\beta\dtau\dtau' K(\tau-\tau')
\cos\pll{\theta^-(\tau)+\theta^-(\tau')\over2}
\prl,
\label{2.31}
\end{equation}
\end{full}
where $K(\tau-\tau')\equiv\int{\rm d}r_\perp{\rm d}r'_\perp K(x,x')$.
From eqs.(\ref{2.21}) and (\ref{2.31}), we obtain
contributions to the action from the electron field as,
\begin{full}
\begin{equation}
S_{\el}={1\over 2\beta}\sum_n\theta_n^- {\alpha|\omega_n|\over2\pi}
\theta_{-n}^-
-\int_0^\beta\dtau\dtau' K\pls\tau-\tau'\prs
\cos\pll{\theta^-\pls \tau\prs+\theta^-\pls \tau'\prs\over2}\prl.
\end{equation}
\end{full}

With the help of eq.(\ref{2.9}), the contribution from $S_{\rm em}$, eq.(\ref{2.4}c), is given as,
\begin{equation}
S_{\rm em}=\int_0^\beta \dtau
{C\over8e^2}\pls{\partial\theta^-\over\partial\tau}\prs^2,
\label{2.32}
\end{equation}
where $C$ is the capacitance of the junction.
In the derivation of eq.(\ref{2.32}), we have assumed that $\theta$ is uniform 
in $r_\perp$ direction and used the approximation,
${\mib \nabla} \theta\sim{\theta^-/L}$.
The total action for the phase is now reduced to the effective action, 
$S_\eff$, which is written with the single degrees of freedom, $\theta$,
as,
\begin{full}
\begin{subeqnarray}
S_\eff&=&{1\over2\beta}\sum_n \theta_n g_{0,n}^{-1} \theta_{-n}
-\int_0^\beta\dtau\dtau'K\pls \tau-\tau'\prs \cos\pll
{\theta\pls
\tau\prs+\theta\pls\tau'\prs\over2}\prl
-\int_0^\beta\dtau{j\over2e}\theta\pls \tau\prs,\ \\
g_{0, n}^{-1}&=&m\omega_n^2+{\alpha\over2\pi}|\omega_n|,\\
m&=&{C\over4e^2}\equiv{1\over2E_c},\\
\alpha&=&{R_Q\over R_N}.
\label{2.35}
\end{subeqnarray}
\end{full}
where we have denoted $\theta^-$ as $\theta$.
The last term of eq.(\ref{2.35}a) is added to the action to express the bias current, $j$.

When the kernel $K(\tau)$ is a delta function,
the effective action, eq.(\ref{2.35}) reduces to that of the
phenomenological model of resistively and capacitively shunted
junction(RCSJ), which is a good description for SNS junction in
the classical limit.

The action, eq.(\ref{2.35}a), differs from that of Josephson
junction in two respect.
First, there is an Ohmic dissipation term, $\alpha$, in eq.(\ref{2.35}b), whereas in 
the action of Josephson junction the dissipation term is given by sinusoidal 
form reflecting the discrete transfer of charge between the electrodes.
Second, the kernels
are short ranged in Josephson junction, whereas, in SNS
junction, the kernel, $K(\tau)$, has a long time tail due to the low energy
excitation of the N region.

The effective action, eq.(\ref{2.35}), is derived for the first time 
in SNS junction. We will compare this action to the similar system of 
SINIS junction, which is derived by Guinea and Sch\"on.\cite{GS,BFS}
Due to the fact that there is a voltage drop across the SIN boundaries,
there are two degrees of freedom, $\theta_R$ and $\theta_L$, which represents
the phase difference across right and left SIN boundaries, respectively.
The contribution of the diagram in Fig.~\ref{diagrams}b leads to the 
dissipative term in this case.
Josephson coupling term, however, leads to 
\begin{equation}
-\int_0^\beta\dtau\dtau' K(\tau-\tau')\cos\pls \theta_R\pls \tau\prs
-\theta_L\pls\tau'\prs\prs.
\label{Guinea}
\end{equation}
Here the kernel $K(\tau-\tau')$ is given by the pair propagator as in
our results, eq.(\ref{2.35}a), but the difference of the 
argument of the cosine term in eq.(\ref{Guinea}) is to be noted.

Finally we comment on the assumption that $\vn \theta$ can be approximated
as a constant in N-region. This means constant electric field is assumed 
in N-region, although the electric field must be determined self-consistently 
at each point in N-region in principle.
Since our aim is to investigate the retardation effect of $K(\tau)$, 
the essential physics will be seen in this approximation. 

In the next section, we will discuss the
imaginary part of the free energy based on the partition function,
\begin{equation}
Z=\int\D\theta\exp \pls -S_\eff\prs.
\end{equation}
Before going into the discussion of tunneling rate,
we will discuss the time dependence of the kernel $K(\tau)$ in the
following sub-section.

\subsection{Time Dependence of the Kernel}
The long time behavior of the kernel $K(\tau)$ is determined by the
low energy excitation of the system.
In Josephson junction, 
$K(\tau)$ can be approximated by $\delta$-function
due to the existence of gap, $\Delta$,
in the excitation spectrum.
In SNS junction, however, time dependence of $K(\tau)$ is determined from 
the low energy excitation of the N region.
Various situation is expected depending on the difference
in the electronic properties.
In this paper, we will consider
the case where the motion of the electrons in 2DEG is diffusive.
Actually, the experiments have been carried out even in the strong
localization regime. We expect, however, the essential aspects of physics
will be clarified by studying such a diffusive case.

The kernel $K(\tau)$ is given in the form,
\begin{equation}
K\pls\tau\prs=E_{J0}{\hat k}(L, \tau).
\end{equation}
Here ${\hat k}(L, \tau)$ is the real space representation of the
pair propagator, which is normalized appropriately.
Since we are interested in the dependence of $K(\tau)$ on diffusion constant,
$D$, the other factors, such as density of states,
transfer integral and so on are included in $E_{J0}$, which 
is given by
\begin{equation}
E_{J0}={4\over \pi}{t^4 m_e^{2}N(0)\over L^2},
\end{equation}
where $N(0)$ is the density of state of 2DEG per spin. 

In the region where the electron motion is diffusive, the
pair propagator, $k(q,{\rm i}\omega_n)$, is given 
for $\omega\tau_\tr \ll 1$ and $q\ll l^{-1}$ as,
\begin{equation}
k(q, {\rm i}\omega_n)={1\over \beta}\sum_{\nu_n>0}^{\nu_c}
{1\over 2 \nu_n+|\omega_n|+D q^2 },
\label{2.43}
\end{equation}
where $\nu_n=(2n+1)\pi T$ and $\nu_c$ is the cutoff frequency.
The diffusion constant is given by
$D=v_F^2\tau_\tr/2=\pi\sigma/(e^2 m_e)={2L\alpha/m_e W}$ with 
$\alpha=R_Q/R_N$.
The function, ${\hat k}(L,\tau)$, is given by Fourier transform
of eq.(\ref{2.43}) with a suitable boundary condition.
By the boundary condition that the current normal to the SN boundary is 
zero\cite{SK,Kresin}, we obtain
\begin{equation}
{\hat k}(L, {\rm i}\omega_n)={1\over4}
\sum_{m=-\infty}^\infty (-1)^m k({\pi\over L}m, {\rm i}\omega_n).
\label{2.45}
\end{equation}
From the  relation, eq.(\ref{2.45}), $K(\tau)$ is given as,
\begin{full}
\begin{subeqnarray}
K(\tau)&=&E_{J0}{1\over\beta}\sum_{n}e^{-{\rm i}\omega_n\tau}{\hat k}
\pls L, i\omega_n\prs\nonumber\\
&=&E_{J0}{\cal P}\int_{-\infty}^\infty{\dw\over\pi}{{\rm e}^{-\omega|\tau|}\over
1-{\rm e}^{-\beta\omega}}
\Im {\hat k}^{(R)}(L, \omega),\\
{\hat k}^{(R)}(L, \omega)&=&{\hat k}(L, \omega+{\rm i}0^+).
\label{2.46}
\end{subeqnarray}
\end{full}

The real time behavior of the kernel, $K({\rm i}t)$,
is derived from $K(\tau)$ by the analytic continuation, {\it i.e.}, 
$\tau\to {\rm i}t$, as,
\begin{full}
\begin{subeqnarray}
K({\rm i}t)&=&K_R(t)+{\rm i}K_I(t),\\
K_R(t)&\equiv&E_{J0}\int_{-\infty}^\infty
{\dw\over\pi}{\cos{\omega t}\over 1-e^{-\beta\omega}}
\Im {\hat k}\pls L, \omega\prs,\\
K_I(t)&\equiv&-E_{J0}\int_{-\infty}^\infty
{\dw\over\pi}{\sin{\omega t}\over 1-e^{-\beta\omega}}
\Im{\hat k}\pls L, \omega\prs.
\end{subeqnarray}
\end{full}

%#### Figure ####
\begin{fullfigure}
\epsfile{file=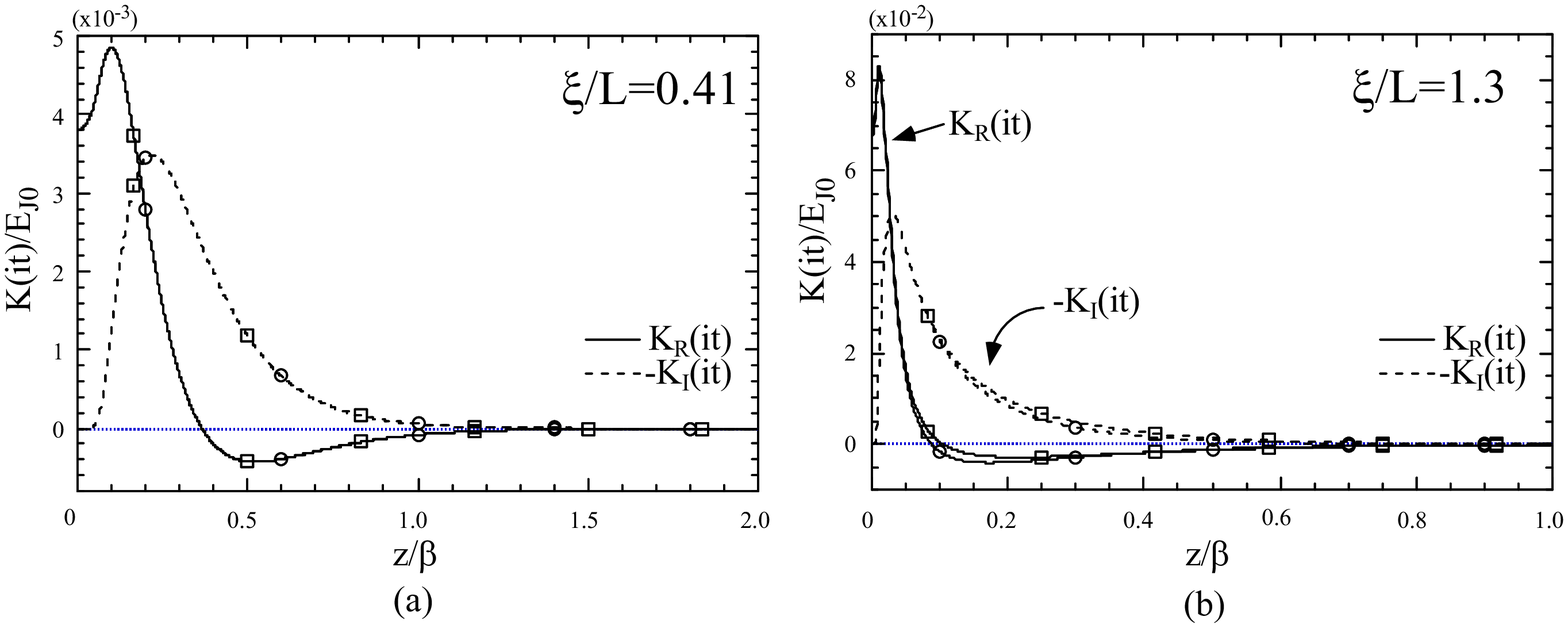,width=14cm}
\caption{The kernel $K_R(t)$(solid line) and $-K_I(t)$(dashed line)
as a function of real time $t/\beta$ for (a) $\xi/L=0.41$ and
(b) $\xi/L=1.3$. 
The lines with square is for the boundary
condition that the current normal to the SN boundary is zero and 
the line with circle represents its $L\to \infty$ limit.}
\label{kernel:time}
\end{fullfigure}
%################

The time dependence of the kernel, $K(i\tau)$, is determined by $\xi/L$, where
$\xi\equiv\sqrt{D/2\pi T}$ is the coherence length. 
The coherence length is related to $\alpha$ as 
$\xi/L=\sqrt{\alpha/\pi m_e WLT}$. 
To make a direct contact with the experiment of Takayanagi {\it et al},~\cite{THN1}
we take $W=80{\rm \mu m}$, $L=0.4{\rm \mu m}$, $T=20 {\rm mK}$ and   
$m_e=0.026m_0$(InAs) and the Nb electrode is used.
Here $m_0$ is the free electron mass. 
With these parameters, $\xi/L$, is given as 
$\xi/L=1.3\times 10^{-2}\sqrt{\alpha}$. 
Thus we obtain $\xi/L=0.41$ for $\alpha=10$ and $\xi/L=1.3$ for $\alpha=100$, 
respectively. 
The time dependence of $K(it)$ for these values are shown in 
Fig.~\ref{kernel:time}.
Here the solid line expresses $K_R$ and the dashed line $K_I$.
The line with square represents the kernel and with the circle represents 
the asymptotic form for $L\to \infty$. 
We take the cutoff frequency $\nu_c$ as $\nu_c\sim \Delta$.
This is consistent with the approximation that the Green's function 
in S-region is a $\delta$ function. 
From the graph, we see little difference for the time dependence between 
asymptotic form and the exact one even for $\xi/L=1.3$.
Thus we will use the limiting form hereafter, for the computational 
simplicity.

\section{Calculation of Tunneling Rate}
In this section, we will explain the method to calculate the tunneling rate 
of the Josephson phase from the metastable state.
In our action, due to the time dependence of the Kernel, 
$K(\tau)$, the potential term and the dissipation term cannot be 
divided easily. 
Thus we employ the self consistent harmonic approximation as 
Geigenm\"uller and Ueda.~\cite{GU}
Applying self-consistent harmonic approximation, we first calculate the 
renormalization of the tunneling potential, the mass and the strength of 
dissipation. 
With these parameters, the decay rate formula for the cosine potential is 
applied to calculate the tunneling rate.

In the next sub-section, we will derive the self-consistent equation.

\subsection{Self Consistent Harmonic Approximation}
In order to separate the potential term and dissipation term in our 
effective action, we use the self consistent harmonic approximation. 
This will be done by using the variational principle, 
which was developed by Feynman and Kleinert.~\cite{FK}

The variational principle for the free energy is given by,\cite{F:text2}
\begin{eqnarray}
F&\le&F_{\rm tr}+{1\over\beta}\langle S-S_{\rm tr}\rangle_{\rm tr},\\
e^{-\beta F_{\rm tr}}&\equiv&\int{\cal D}\theta e^{-S_{\rm tr}}.
\label{3.1}
\end{eqnarray}
Here $\langle\cdots\rangle_{\rm tr}$ denotes the average with respect to 
trial action $S_{\rm tr}$.
This follows from the convexity of the Boltzmann weight function with 
respect to the action.

To determine the renormalization of potential, mass and strength of 
dissipation, 
the direct application of eq.(\ref{3.1}) is not possible. 
Thus we will define the effective classical potential, $W(\bar{\theta})$, and 
consider the motion of the phase in $W(\bar{\theta})$. 
The effective classical potential is defined as,\cite{FK,We}
\begin{equation}
\exp(-\beta W(\bar{\theta}))\equiv \int {\cal D}\bar{\Theta}(\tau) 
\exp\pls -S_{\rm eff}\prs,  
\label{3.2}
\end{equation}
where $\bar{\Theta}(\tau)$ is the fluctuation around the average path,
$\bar{\theta}=\int_0^\beta \dtau \theta\pls\tau\prs/\beta$, which is defined 
by 
\begin{eqnarray}
\bar{\Theta}(\tau)&\equiv&\theta(\tau)-\bar{\theta},\nonumber\\
&=&{1\over\beta}\sum_{n\neq0}e^{-{\rm i}\omega_n\tau}\theta_n.
\end{eqnarray}
Here Matsubara frequency is given by $\omega_n=2\pi n/\beta$.
With the effective classical potential, the partition function is 
expressed as,
\begin{equation}
Z=\int {\rm d}\bar{\theta} e^{-\beta W(\bar{\theta})}.
\label{3.4}
\end{equation}
This form resembles that of the classical partition function with potential 
W. This is the reason we call W, the effective classical potential. 
This is the exact limiting form of the partition function for high temperature,
as the deviation of the path from $\bar{\theta}$ is small in this case. 

In a damped harmonic oscillator, we can perform the path
integral in eq.(\ref{3.2}) exactly, since all the
integrals are quadratic in the variable $\theta_n$.
In our case, however, the path integration cannot be carried out, 
and some kind of approximation is needed. Here we apply the
variational principle for W.
Since $\exp(-\beta W)$ is convex as a function of $W$, the same 
discussion as the derivation of variational principle for the free 
energy can be applied. The variational principle for $W$ is given by, 
\begin{equation}
W\leq W_{\tr'}+{1\over\beta}\langle S-S_{\tr'}\rangle_{\tr'}
\equiv{\tilde W},
\label{VP}
\end{equation}
where $W_{\tr'}$ is a trial effective classical potential evaluated with 
$S_{\tr'}$.Here prime denotes that $\bar{\theta}$ degree of freedom is 
excluded. 

Above discussion holds, as far as the function,
$\exp\pll{-\beta W\pls\bar{\theta}\prs}\prl$, is real.
However, when the bias current $j$ comes closer to the critical current, 
the variational parameters acquire imaginary parts 
due to the fact that the potential barrier height becomes comparable to 
the zero point energy of the quasi ground state in the potential well.
In such situation, the inequality, eq.(\ref{VP}), no longer holds.
Nevertheless, based on the discussion of Kleinert,
we expect that the best approximation to the effective classical 
potential is obtained when the derivative of $\tilde W$ with respect to 
$(S-S_{\tr'})$ equals to zero.
\cite{Kleinert1}
The reason is as follows.
A full perturbation expansion of $\exp(-\beta W)$ with respect to $S_{\tr'}$ 
is given as,
\begin{full}
\begin{eqnarray}
W(\bar{\theta})=-{1\over\beta}\ln Z_{\tr'}
+{1\over\beta}\langle \pls S-S_{\tr'}\prs\rangle_{\tr',c}
+{1\over2\beta}\langle \pls S-S_{\tr'}\prs^2\rangle_{\tr',c}
+\cdots.
\end{eqnarray}
\end{full}
Here $\langle\cdots\rangle_{\tr', c}$ denotes the cumulant average.
By definition, above expansion is clearly independent of the trial
action, $S_{\tr'}$, if we sum the series to infinite order. 
This independence, however, is lost when we truncate the series to 
the finite order. 
But we expect the best trial action when the approximation shows minimal 
dependence on $S_{\tr'}$.
Thus even in the region where the parameters have imaginary parts, 
we shall use the same variational free energy as in the region where the 
parameters have no imaginary parts.

Now let us apply eq.(\ref{VP}) to the effective action
of the SNS junction given by eq.(\ref{2.35}).
We take the trial action of the form,
\begin{subeqnarray}
S_\tr&\equiv&
{1\over2\beta}\sum_{n\neq0}\theta_n g_n^{-1} \theta_{-n},\\
g_n^{-1}&\equiv&m_{\rm ren}\omega_n^2+{\alpha_{\rm ren}\over2\pi}\al\omega_n\ar+m_{\rm ren}
\Omega_{\rm ren}^2\nonumber\\
&\equiv&g^{-1}({\rm i}\omega_n).
\label{3.16}
\end{subeqnarray}
With this trial action, we obtain for $\tilde W$,
\begin{full}
\begin{equation}
{\tilde W}={1\over\beta}\log\prod_{n=1}^\infty{1\over m\omega_n^2g_n}
+{1\over2\beta}\sum_{n\neq0}\pls{g_n\over g_{0n}}-1\prs\nonumber\\
-{j\over2e}\bar{\theta}_0-\int_0^\beta\dtau K(\tau)e^{-{1\over4}
\pl g(0)+g(\tau)\pr }
\cos\bar{\theta},
\end{equation}
\end{full}
where $g(\tau)=\sum_n e^{-{\rm i}\omega_n\tau}g_n/\beta$.
Taking the functional derivative of $\tilde W$ with respect to $g_n$, 
we obtain,
\begin{full}
\begin{eqnarray}
\delta{\tilde W}&=&{1\over\beta}\sum_{n=1}^\infty
\delta g_n\pll
-{1\over g_n}+{1\over g_{0n}}\pre+\ple{1\over2}\int_0^\beta\dtau
K\pls \tau\prs \pls 1+e^{-{\rm i}\omega_n\tau}\prs
e^{-{1\over4}\pl g\pls0\prs+g\pls\tau\prs\pr}
\cos\bar{\theta}\prl\nonumber\\
&=&0\label{3.22.4}.
\end{eqnarray}
\end{full}
Next, to find out the potential minima,$\bar{\theta}_0$, we differentiate
$\tilde{W}$ with respect to $\bar{\theta}$,
which leads to,
\begin{full}
\begin{eqnarray}
{\delta {\tilde W}\over\delta\bar{\theta}}
&=&\int_0^\beta\dtau K\pls\tau\prs
e^{-{1\over4}\pl g\pls0\prs+g\pls\tau\prs\pr}
\sin\bar{\theta}-{j\over2e}
-{1\over\beta}\sum_{n=1}^\infty
{\delta g_n\over\delta\bar{\theta}}
\pll
g_n^{-1}-g_{0n}^{-1}
-{1\over2}\int_0^\beta
K\pls\tau\prs
e^{-{1\over4}\pl g\pls0\prs+g\pls\tau\prs\pr}
\cos\bar{\theta}\prl\nonumber\\
&=&\int_0^\beta\dtau K\pls \tau\prs e^{-{1\over4}
\pl g\pls 0\prs+g\pls\tau\prs\pr}
\sin\bar{\theta}-{j\over2e}\nonumber\\
&=&0.\label{3.24}
\end{eqnarray}
\end{full}
Here we have used the variational condition, eq.(\ref{3.22.4}), in the second 
equality.
Above conditions, eqs.(\ref{3.22.4}) and (\ref{3.24}), lead to a set of 
self-consistent equations,
\begin{full}
\begin{subeqnarray}
g_n^{-1}-g_{0n}^{-1}&=&{1\over2}\int_0^\beta\dtau
K(\tau)(1+e^{-{\rm i}\omega_n\tau})
e^{-{1\over4}\pl g(0)+g(\tau)\pr}\cos\bar{\theta}_0,\\
{j\over2e}&=&\int_0^\beta\dtau K\pls\tau\prs
e^{-{1\over4}\pl g(0)+g(\tau)\pr}\sin\bar{\theta}_0.
\label{sceq}
\end{subeqnarray}
\end{full}
These equations are formally the same as those derived by
Geigenm\"uller and Ueda in Josephson junction, when we replace $K(\tau)$ by 
corresponding terms 
in the action which has been obtained by Ambegaokar et al.~\cite{GU}
However, their approach differs from ours in the following point:
They took $\bar{\theta}_0$ as an independent variable, whereas in
our approach $j$ is the independent variable. For this reason,
they can not discuss the region where the imaginary parts appear 
in the renormalized parameters. Furthermore, by noticing that 
the bias current $j$ is the 
controllable variable in experiments, our choice seems more natural.

\begin{figure}
\epsfile{file=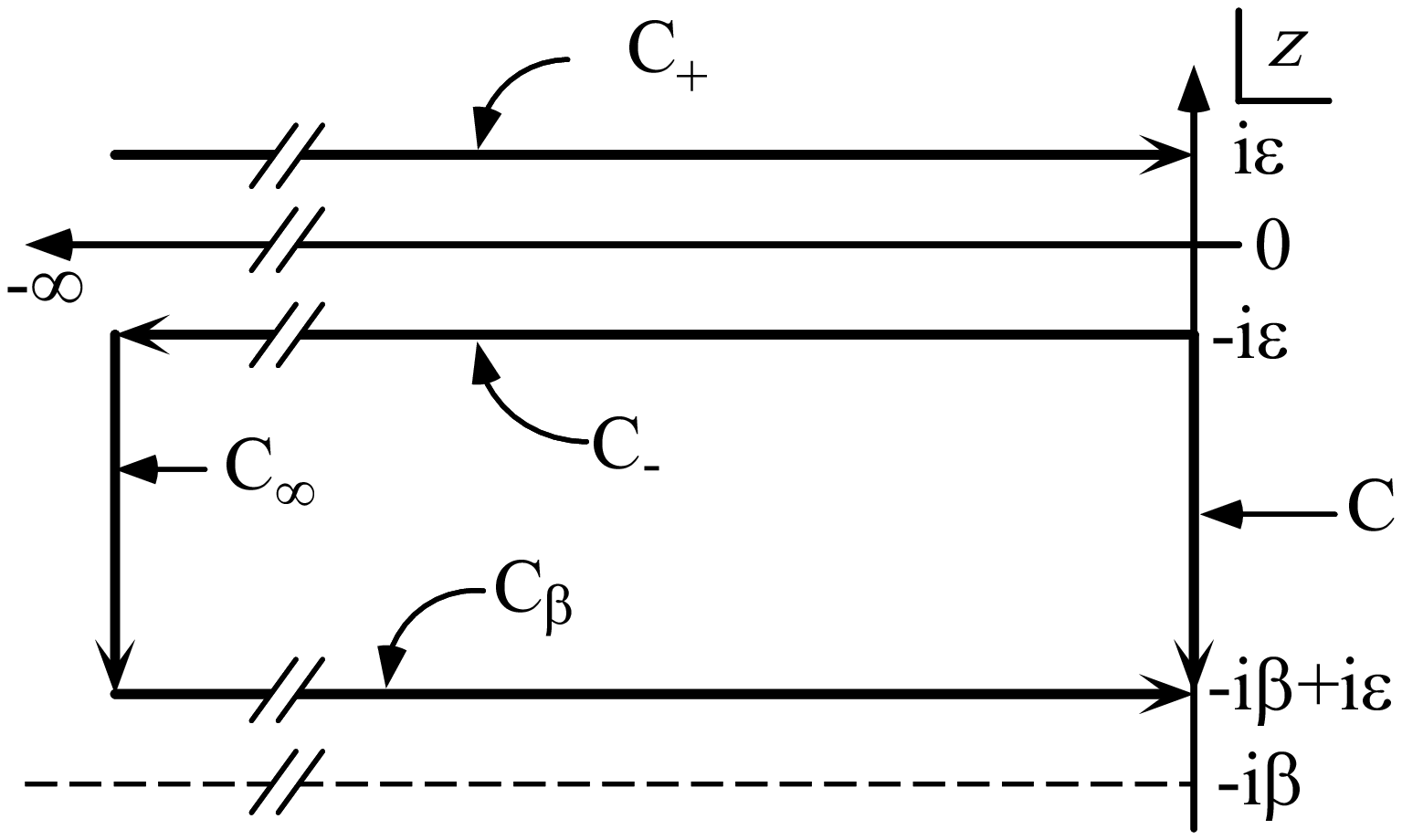,width= 7cm}
\caption{Integration contour used when deriving eqs.(3.14) 
from eq.(3.13).}
\label{IC}
\end{figure}

Our aim is to calculate the tunneling rate of the phase.
For that purpose, we have to determine the renormalized mass,
$m_{\rm ren}$, renormalized strength of the dissipation,
$\alpha_{\rm ren}$, and renormalized attempt frequency,
$\Omega_{\rm ren}$, which are defined at low energy.
To get these variables, we have to perform analytic continuation of
the Green's function in eq.(\ref{sceq}) from imaginary frequency 
to real frequency.
We first transform the variables by $\tau={\rm i}z$ and
change the integration contour as shown in Fig.~\ref{IC}.
Next we perform analytic continuation, 
${\rm i}\omega_n\to\omega+{\rm i}0^+$, 
and expand the resulting equation with respect to $\omega$.
Thus we obtain the following four sets of equations as shown \
in the appendix,~\cite{AWAK}
\begin{full}
\begin{subeqnarray}
m_{\rm ren}&=&m+{1\over4}e^{-{1\over4}g\pls 0 \prs}
\int_0^\infty{\rm d}z z^2 e^{-{1\over4}g_R\pls z\prs}
\pll K_R\pls z \prs \sin{g_I\pls z \prs\over4}
-K_I\pls z\prs\cos{g_I\pls z\prs\over 4}\prl
\cos\bar{\theta}_0,\ \ \ \ \ \ \ \ \\
{\alpha_{\rm ren}\over2\pi}
&=&{\alpha\over2\pi}-{1\over2}e^{-{1\over4}g\pls 0 \prs}
\int_0^\infty{\rm d}z z e^{-{1\over4}g_R\pls z\prs}
\pll K_R\pls z\prs \sin{g_I\pls z \prs\over4}
-K_I\pls z\prs\cos{g_I\pls z\prs\over 4}\prl
\cos\bar{\theta}_0,\\
m_{\rm ren}\Omega_{\rm ren}^2&=&e^{-{1\over4}g\pls 0 \prs}
\int_0^\infty{\rm d}z e^{-{1\over4}g_R\pls z\prs}
\pll K_R\pls z\prs \sin{g_I\pls z \prs\over4}
-K_I\pls z\prs\cos{g_I\pls z\prs\over 4}\prl
\cos\bar{\theta}_0,\\
{j\over2e}&=&e^{-{1\over4}g\pls 0 \prs}
\int_0^\infty{\rm d}z e^{-{1\over4}g_R\pls z\prs}\pll K_R\pls z
\prs \sin{g_I\pls z \prs\over4}
-K_I\pls z\prs\cos{g_I\pls z\prs\over 4}\prl \sin\bar
{\theta}_0.
\label{sceq1}
\end{subeqnarray}
\end{full}
Here the Green's functions are given as,
\begin{full}
\begin{subeqnarray}
g\pls {\rm i}z+\epsilon\prs&\equiv&g_R\pls z\prs+{\rm i} g_I\pls z\prs,\\
g_R\pls z\prs&\equiv&
{\cal P}\int_{-\infty}^\infty {\dw\over2\pi {\rm i}}
{\cos \omega z\over 1-e^{-\beta\omega}}
\pll
g^{(R)}\pls \omega\prs
-g^{(A)}\pls \omega\prs
\prl,\\
g_I\pls z\prs&\equiv&
-{\cal P}\int_{-\infty}^\infty {\dw\over2\pi {\rm i}}
{\sin \omega z\over 1-e^{-\beta\omega}}
\pll
g^{(R)}\pls \omega\prs
-g^{(A)}\pls \omega\prs
\prl,
\end{subeqnarray}
\end{full}
where $g^{(R)}(\omega)\equiv g(\omega+{\rm i}0^+)$ and 
$g^{(A)}(\omega)\equiv g(\omega-{\rm i}0^+)$.
%#### Figure ####
%################

In the next chapter, we will solve these equations and calculate the tunneling 
rate at absolute zero.
However, since the integral of kernel $\int_0^\beta \dtau K(\tau)$ 
in eq.(\ref{2.46}) is logarithmically divergent at absolute zero, we will 
use finite temperature form for $K({\rm i}z)$ in the calculation.
  
\subsection{Decay Rate Formula}
As far as the renormalized parameters are real, we
can make use of the Korshnov's formula for the tunneling rate of 
the phase in the large dissipation limit~\cite{Ko}, 
and this condition is achieved in the
neighborhood of the critical current.
We find for the tunneling rate, $\Gamma$,
\begin{full}
\begin{subeqnarray}
\Gamma&=&f_{\rm qm}\exp\pls-S_B\prs,\\
S_B&=&-2\alpha_{\rm ren}\log\sin\bar{\theta}_0,
\\
f_{\rm qm}&\equiv&{j\over2e}\sqrt{{4\pi\over\alpha_{\rm ren}}}
\pls{1\over\sin^2\bar{\theta}_0}-1\prs
\pls{\Delta_1+2}\over{\Delta_1-2}\prs^{\Delta_1},\\
\Delta_1&\equiv&
{2\over\sqrt{1-16\pi^2m_{\rm ren}^2\Omega_{\rm ren}^2/\alpha_{\rm ren}^2}}.
\end{subeqnarray}
\end{full}
Here $\alpha_{\rm ren}$, $m_{\rm ren}$, $\Omega_{\rm ren}$ and $\theta_0$ 
are determined from eq.(\ref{sceq1})

%#### Figure ####
\begin{figure}
\epsfile{file=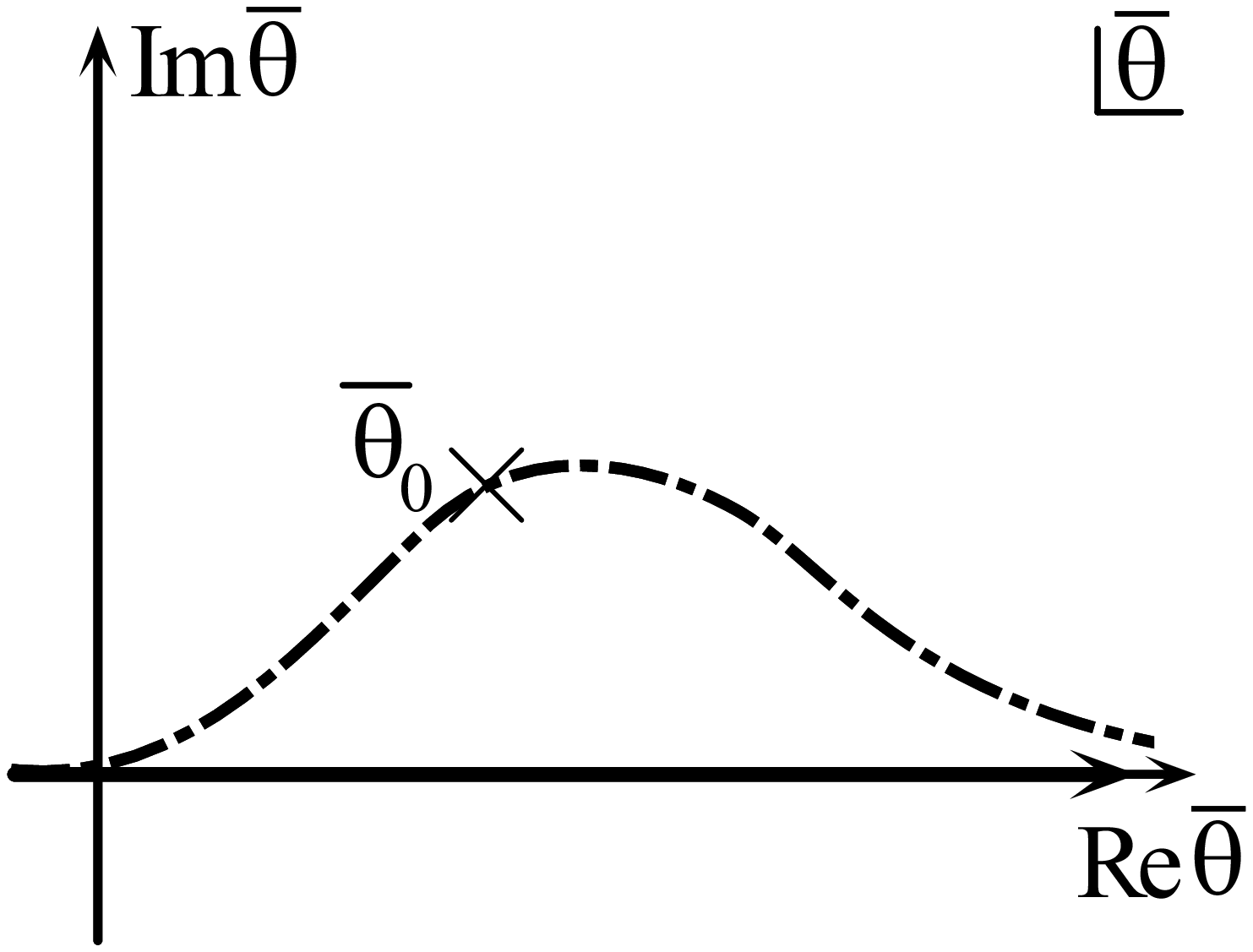,width= 7cm}
\caption{The path used in the saddle point approximation of eq.(3.17)(dot-dashed 
line).
The saddle point is $\bar{\theta}_0$.}
\label{DP}
\end{figure}
%################

On the other hand, when the renormalized parameters acquire
imaginary part, the free energy obtains imaginary part correspondingly.
The appearance of the imaginary part reflects the fact
that the potential well is no longer metastable, but rather the phase
can decay directly without the barrier.
Hence we can regard this imaginary part of the free energy
as the direct decay rate of the phase.
The calculation of the imaginary part is performed
by saddle point approximation.
In this case the partition function will be estimated by changing 
the contour as in Fig.~\ref{DP},
\begin{full}
\begin{eqnarray}
Z&\equiv&e^{-\beta F}\nonumber\\
&=&\int{{\rm d}\bar{\theta}\over\sqrt{2\pi\beta/m}}
e^{-\beta W\pls\bar{\theta}\prs}\nonumber\\
&\sim& e^{-\beta W\pls\bar{\theta_0}\prs}
\int_{-\epsilon e^{-{\rm i}\zeta/2}}^{\epsilon e^{-{\rm i}\zeta/2}}
{{\rm d}y\over\sqrt{2\pi\beta/m}}
e^{-{\beta\over2}W''\pls\bar{\theta_0}\prs y^2},
\end{eqnarray}
\end{full}
where $\zeta$ is the argument of $W(\bar{\theta}_0)$.
From this we obtain for the imaginary part of free energy as,
\begin{equation}
\Im F=\Im W(\bar{\theta}_0)+{\zeta\over\beta}.
\label{3.37-}
\end{equation}
Here $\Im W$ is given by following formulae,~\cite{AWAK}
\begin{full}
\begin{subeqnarray}
\Im W \pls\bar{\theta}_0\prs
&=&\Im\pll W_1+W_2+W_3+W_4\prl,\\
\Im W_1&=&{\omega_c\over2\pi}\Im\pll\log{m_{\rm ren}\over m}
+\log\pls 1+{\gamma_{\rm ren}\over\omega_c}
+{\Omega_{\rm ren}^2\over\omega_c^2}\prs\prl-\Im\pll {\omega_+\over2\pi}
\log{\omega_c-\omega_+\over-\omega_+}
+{\omega_-\over2\pi}
\log{\omega_c-\omega_-\over-\omega_-}\prl,\ \ \ \ \ \ \ \ \ \\
\Im W_2&=&{\omega_c\over2\pi}\Im{m\over m_{\rm ren}}
+{1\over2\pi}\Im{m\over m_{\rm ren}}{\gamma-\gamma_{\rm ren}
-\Omega_{\rm ren}^2\over\omega_+-\omega_-}
\pll {\omega_+\over2\pi}\log{\omega_c-\omega_+\over
-\omega_+}
-{\omega_-\over2\pi}\log{\omega_c-\omega_-\over-\omega_-}
\prl,\\
\Im W_3&=&-{j\over2e}\Im\bar{\theta}_0,\\
\Im W_4&=&-\Im {m_{\rm ren}\Omega_{\rm ren}^2}.
\label{3.37}
\end{subeqnarray}
\end{full}
Here $\omega_c$ is the cut-off frequency,
$\gamma_{\rm ren}$, $\omega_+$ and $\omega_-$ are given by,
\begin{subeqnarray}
\gamma_{\rm ren}&\equiv&{\alpha_{\rm ren}\over2\pi m_{\rm ren}},\\
\omega_\pm&\equiv&{\gamma_{\rm ren}\over2}\pll-1\pm\Delta_2\prl,\\
\Delta_2&=&\sqrt{1-{4\Omega_{\rm ren}^2\over\gamma_{\rm ren}^2}}.
\end{subeqnarray}
Decay rate is obtained by the relation $\Gamma=2\Im F$ from eq.(\ref{3.37-}).

In the next section, we will apply these formula to
the effective action, eq.(\ref{2.35}), and investigate
the decay rate of the phase in SNS junction.

\section{Results and Discussion}
In this section, we will explain the results obtained from the effective
action, eq.(\ref{2.35}), by the approximation presented so far.
For the numerical calculation, we have taken $L=0.4{\rm \mu m}$, 
$W=80{\rm \mu m}$ and $C=200{\rm fF}$. 

\subsection{Phase Diagram for $j=0$}
\begin{figure}
\epsfile{file=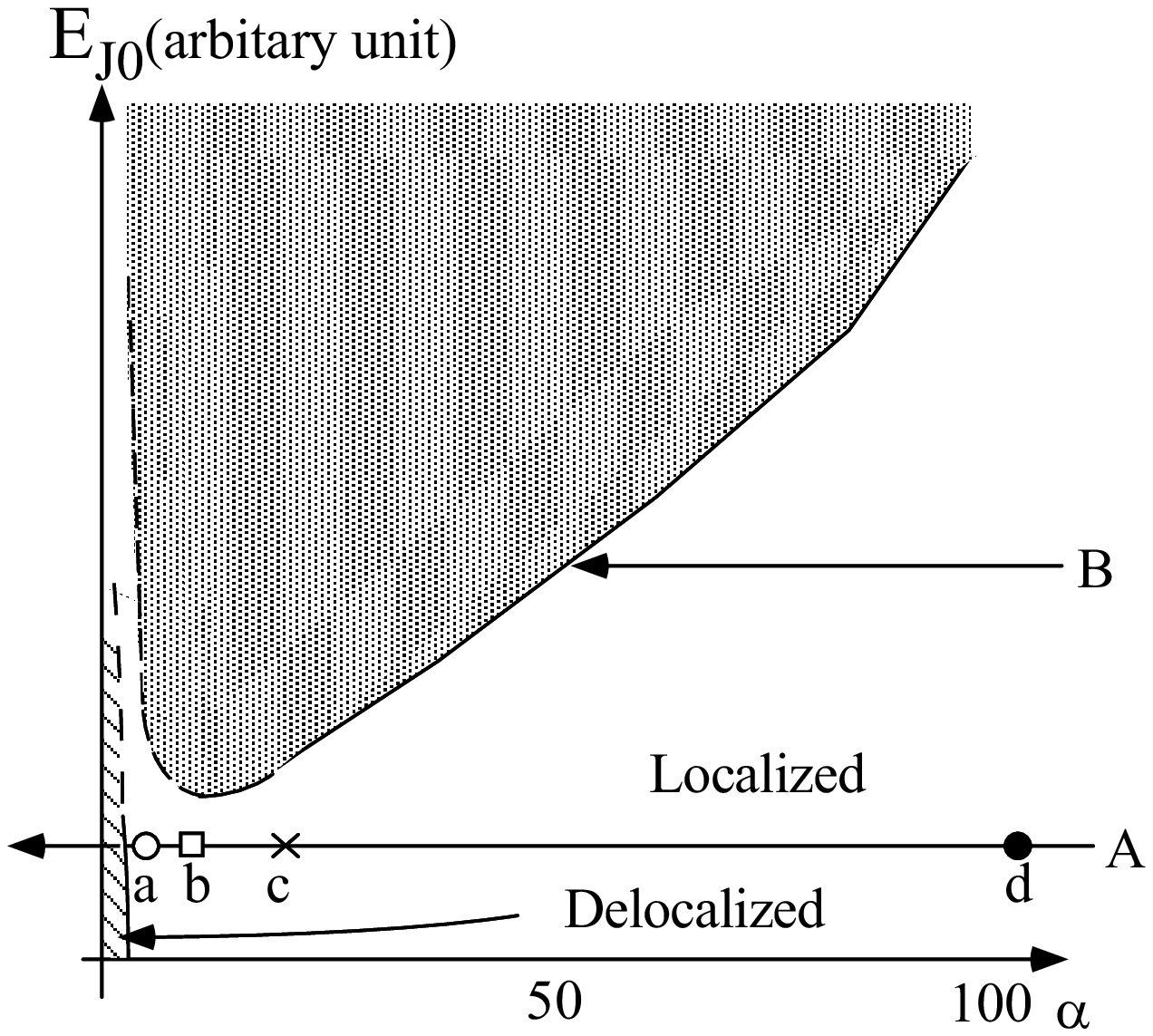,width=7cm}
\caption{The schematic representation of the phase diagram for $j=0$
plotted in the plane of $E_{J0}$ and $\alpha$.
The hatched region is for the delocalized phase and white region  for the
localized phase. In the shaded region, our approximation fails.
}
\label{phase}
\end{figure}

We discuss the case $j=0$ in this subsection.
A schematic representation of the phase diagram in $E_{J0}$-$\alpha$ plane is
shown in Fig.~\ref{phase}.
The white region is for the localized phase($\Omega_{\rm ren}\neq0$),
where the phase variable is trapped in one of the local minima of
the cosine potential.
The hatched region, on the other hand, expresses the delocalized phase($\Omega_{\rm ren}=0$).
In the shaded region, our effective action fails, since we have an 
unphysical result, $\alpha_{\rm ren}<0$. 
The reason for the failure of our action in this region 
will be discussed in subsection 4.3.

%#### Figure ####
%################

Two trivial limiting cases exist in eq.(\ref{sceq1}),
{\it i.e.}, $E_{J0}=0$ and $\alpha=0$.
In both cases, we have  $K({\rm i}z)=0$, 
which leads to the unrenormalized result, 
$m_{\rm ren}=m$, $\alpha_{\rm ren}=\alpha$ and
$\Omega_{\rm ren}=0$. This means the phase is delocalized.
The delocalized phase extends to the finite area for finite values of 
$\alpha$ and $E_{J0}$.
The boundary between the localized and delocalized region lies around
$\alpha\sim1$, when $E_{J0}$ is small and comes closer to $\alpha=0$
as $E_{J0}$ increases.
This is due to the suppression of quantum fluctuation 
by the potential barrier.

For larger $\alpha$, the phase is localized in one of the minimum 
of the cosine well. 
In this region, the effect of cosine term is to
increase $m_{\rm ren}$ and to decrease $\alpha_{\rm ren}$.
In particular, on increasing $E_{J0}$, one observes that $\alpha_{\rm ren}$ 
decreases and finally becomes $0$ and eventually negative.
We can not treat the region with larger $E_{J0}$.(See section 4.3.)

\begin{figure}
\epsfile{file=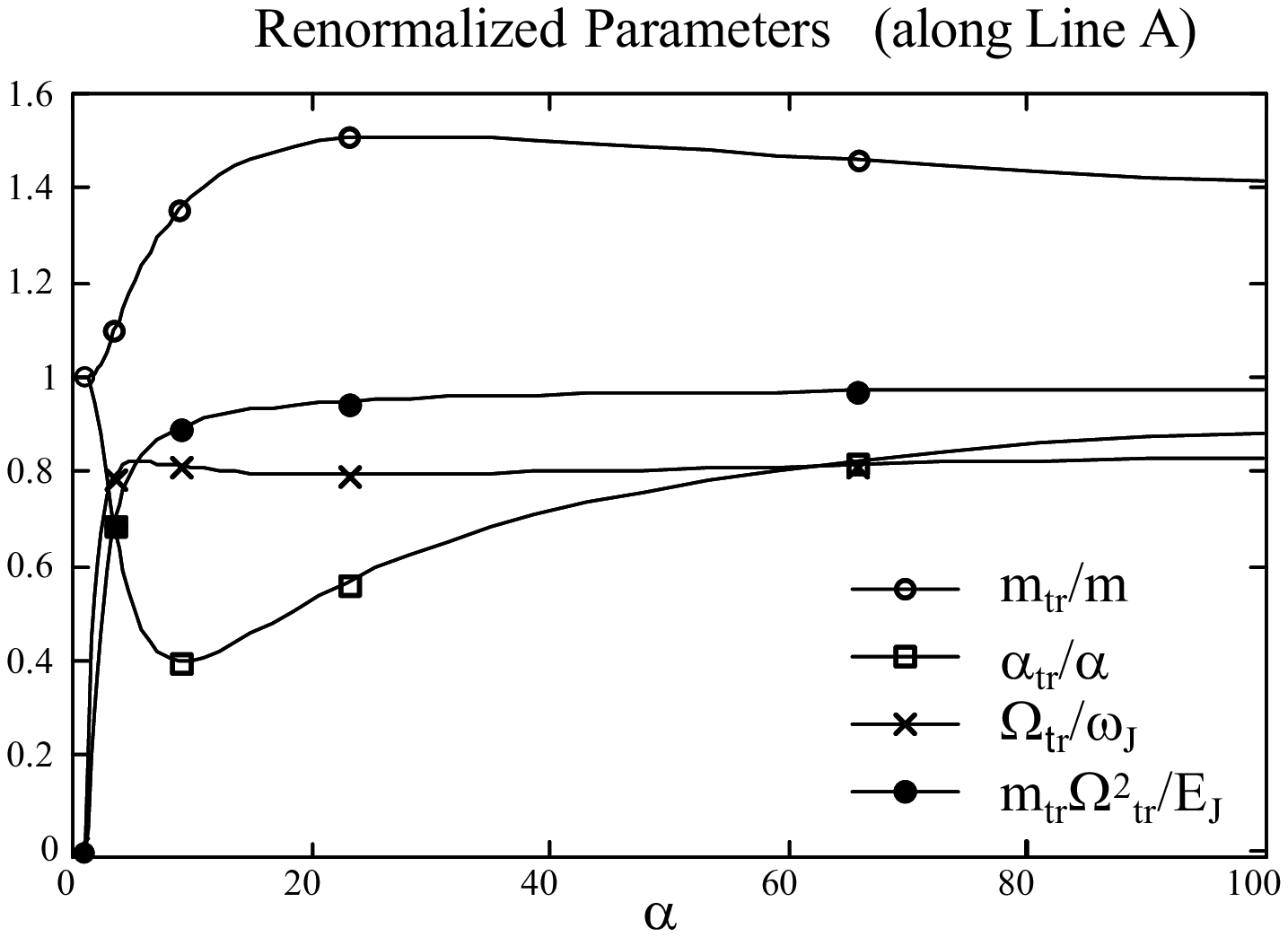,width=7cm}
\caption{Renormalized parameters $m_{\rm ren}/m$, $\alpha_{\rm ren}/\alpha$,
$\Omega_{\rm ren}/\omega_J$ and $m_{\rm ren}\Omega_{\rm ren}^2/E_J$ are plotted as a
function of $\alpha$ along line A.}
\label{LineA}
\end{figure}

We will show the $\alpha$-dependence of renormalized parameters 
along lines A and B in the following.
Those along line A are shown in Fig.~\ref{LineA}.
In this figure, $m_{\rm ren}$, $\alpha_{\rm ren}$, $\Omega_{\rm ren}$ and 
$m_{\rm ren}\Omega_{\rm ren}^2$
are scaled with unrenormalized value, {\it i.e.}, $m$, $\alpha$, $\omega_J$ 
and $m\omega_J^2=E_J$, respectively.
Here Josephson frequency, $\omega_J$, is defined as
$\omega_J=\sqrt{2E_C E_J}$ and $E_J\equiv\int_0^\beta\dtau K(\tau)$.

%#### Figure ####
%################

The behavior observed in Fig.~\ref{LineA} can be understood as follows.
In large $\alpha$ limit, $K({\rm i}z)$ becomes short ranged, as was explained
in chapter 2, and the retardation effect will not appear in the
renormalization of $m_{\rm ren}$, $\alpha_{\rm ren}$ and $\Omega_{\rm ren}$.
Thus, in this limit, $m_{\rm ren}$, $\alpha_{\rm ren}$ and $\Omega_{\rm ren}$ approach to
$m$, $\alpha$ and $\omega_J$, respectively.
On the other hand, in small $\alpha$ limit, the quantum fluctuation of the
phase becomes large and it smears out the effect of cosine term,
and we obtain $m_{\rm ren}\to m$, $\alpha_{\rm ren}\to\alpha$ and 
$\Omega_{\rm ren}\to0$.
This leads to the localization-delocalization transition as a function of
$\alpha$, which was
also found in a system of a particle moving in a periodic potential with
Ohmic dissipation.\cite{Sc,FZ}
For general values of $\alpha$, the effect of cosine term is to increase
$m_{\rm ren}$ and decrease $\alpha_{\rm ren}$.
This tendency can be explained as follows.
By considering the small fluctuation of the phase around $\theta=0$,
we can expand cosine term in the action, eq.(\ref{2.29}), as,
\begin{eqnarray}
S_J&=&-\int_0^\beta\dtau\dtau'K(\tau-\tau')\cos\pls
{\theta(\tau)+\theta(\tau')\over2}\prs\nonumber\\
&\sim&-\int_0^\beta{\rm d}\tau{\rm d}\tau' K(\tau-\tau')\left[1-\left(
{\theta(\tau)+\theta(\tau')\over2}\right)^2\right]\nonumber\\
&\sim&{1\over4\beta}\sum_n\theta_n[K(i\omega_n=0)+K(i\omega_n)]\theta_{-n},
\end{eqnarray}
where we have neglected the constant term. By the analytic
continuation, ${\rm i}\omega_n\to\omega+{\rm i}0^+$, we
find the correction to $m$ and $\alpha$ by the kernel are given by
the low energy behavior of $\Re \pll K^{(R)}(\omega)-K^{(R)}(0)\prl<0$,
and $\Im K^{(R)}(\omega)>0$, respectively.
The signs of these terms account for the enhancement
of $m_{\rm ren}$ and reduction of $\alpha_{\rm ren}$.

\begin{figure}
\epsfile{file=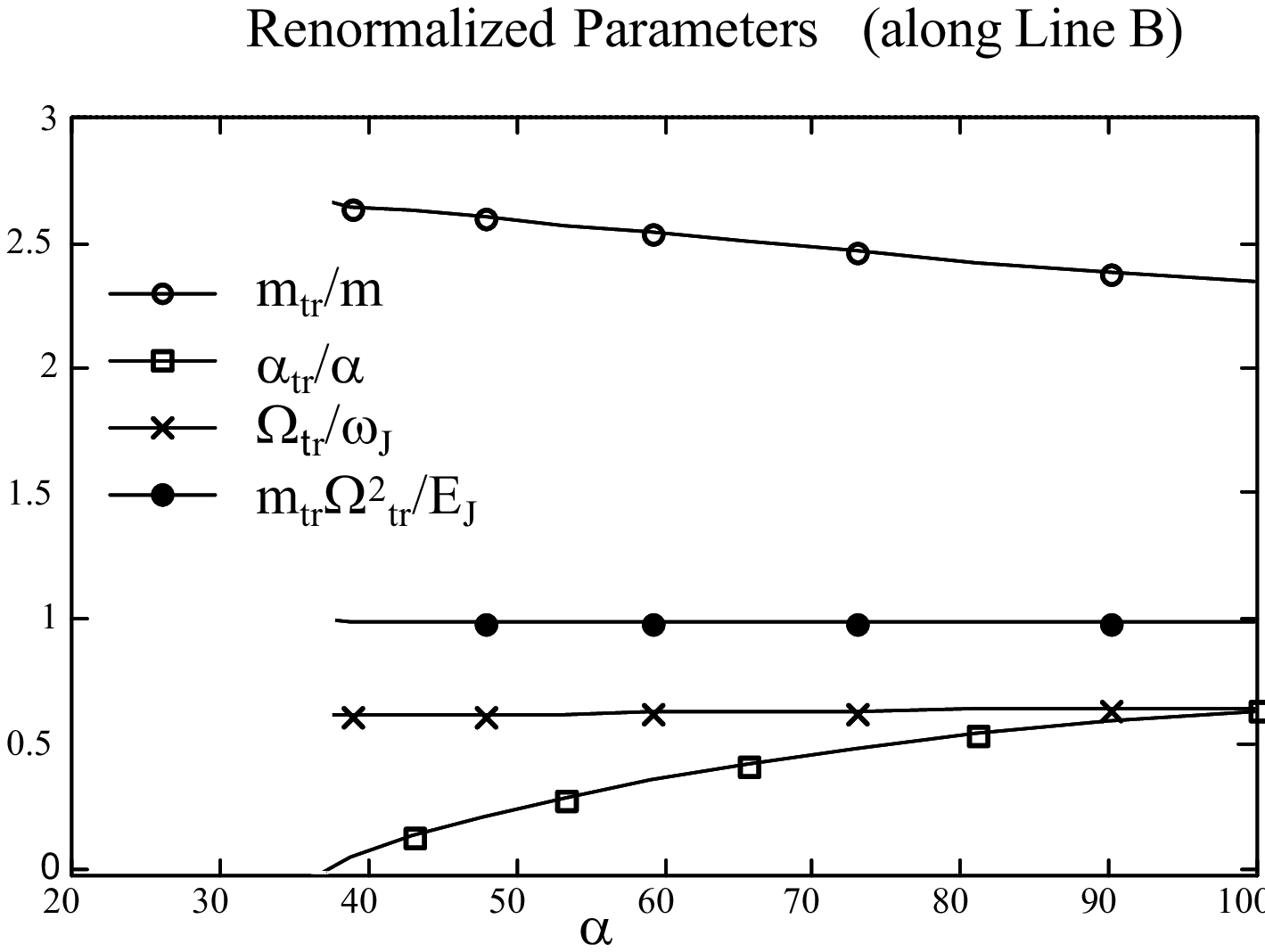,width=7cm}
\caption{Renormalized parameters $m_{\rm ren}/m$, $\alpha_{\rm ren}/\alpha$,
$\Omega_{\rm ren}/\omega_J$ and $m_{\rm ren}\Omega_{\rm ren}^2/E_J$ are plotted as a
function of $\alpha$ along line B.}
\label{LineB}
\end{figure}

Next we will show in Fig.~\ref{LineB} the $\alpha$ dependence of the 
parameters on line B.
For large $\alpha$, the behavior observed on this line is essentially
the same as that observed on line A.
In this case, however, as $\alpha$ is decreased,
 we observe that $\alpha_{\rm ren}$ becomes smaller and finally vanishes.
For smaller $\alpha$, we can not use our effective action.(See section 4.3.)

%#### Figure ####
%################

\subsection{Bias Current Dependence of Renormalized Parameters and Decay Rate}
In this section, we will introduce a finite bias current $j$ and study
the $j$-dependence of variational parameters and the decay rate of
the phase from the metastable state.

%#### Figure ####
\begin{fullfigure}
\epsfile{file=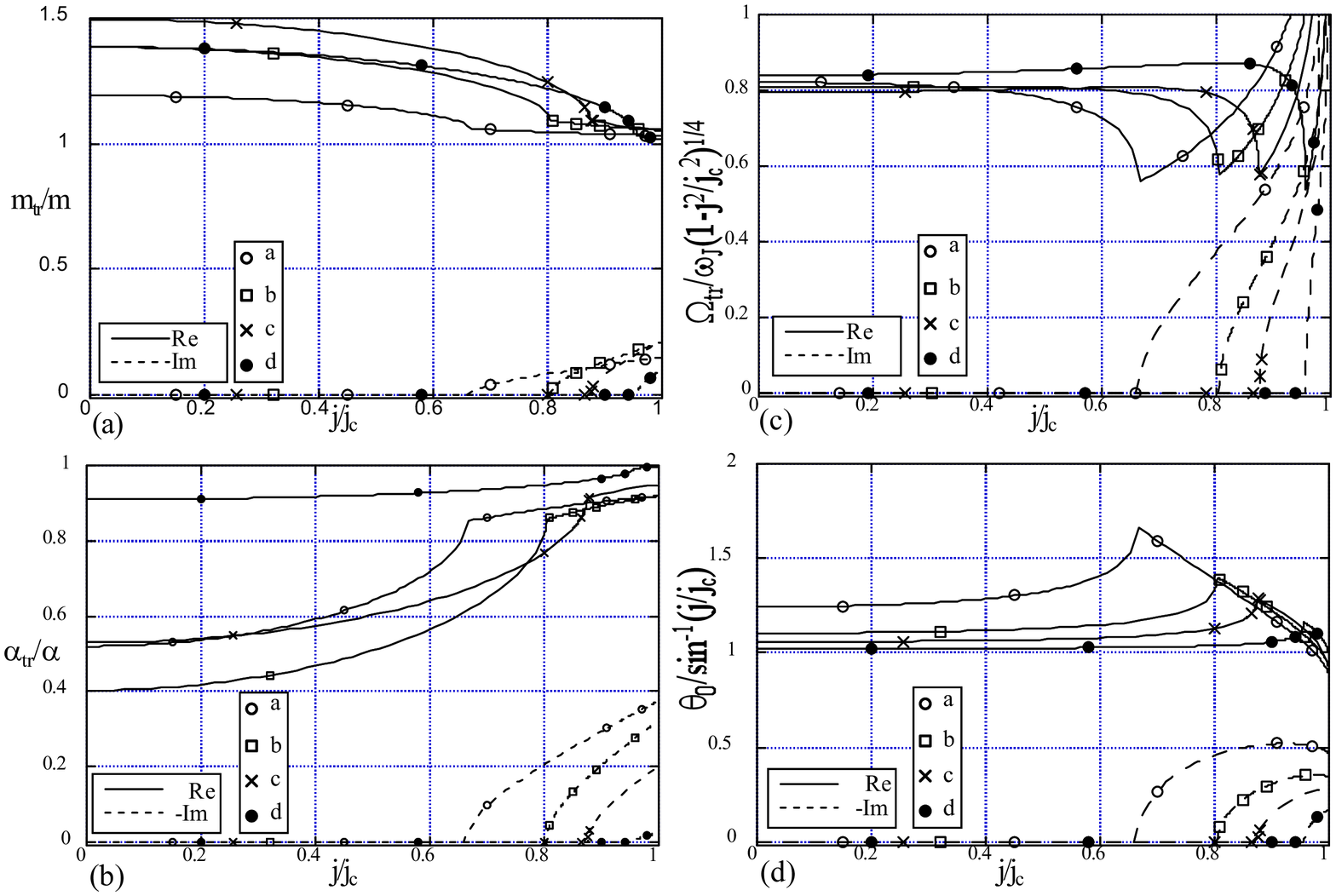,width=14cm}
\caption{Renormalized parameters (a) $m_{\rm ren}/m$, (b) $\alpha_{\rm ren}/\alpha$,
(c) $\Omega_{\rm ren}/\omega_J(1-(j/j_c)^2)^{1/4}$ and 
(d) $\bar{\theta}_0/\sin^{-1}(j/j_c)$
 plotted as a function of bias current $j$, for the points 
(a)$\sim$(d) of Fig. 7.}
\label{param1A}
\end{fullfigure}
%################

In Fig.~\ref{param1A}, $j$ dependences of renormalized mass, $m_{\rm ren}$, 
dissipation,
$\alpha_{\rm ren}$, frequency, $\Omega_{\rm ren}$ and the position of the minimum,
$\bar{\theta}_0$, are shown.
The parameters for these figures are $E_{J0}=1.0{\rm K}$ for all graphs,
and $\alpha=5$ (a), $\alpha=10$ (b), $\alpha=20$ (c) and $\alpha=120$ (d).
The parameters $m_{\rm ren}$, $\alpha_{\rm ren}$, $\Omega_{\rm ren}$ and 
$\bar{\theta}_0$ are
compared with the unrenormalized values, $m$, $\alpha$,
$\omega_J(1-(j/j_c)^2)^{1/4}$ and $\sin^{-1} (j/j_c)$, respectively.
Here $j_c$ is the classical critical current given by $j_c\equiv2e E_J$.
We notice that increasing the current, the renormalization of
$m_{\rm ren}$ and $\alpha_{\rm ren}$ becomes smaller.
This is because the factor $\cos\bar{\theta}_0$ in eq.(\ref{sceq1})
decreases as $j$ is increased and also the quantum
fluctuation reduces the effect of cosine term.
Therefore, $m_{\rm ren}$ and $\alpha_{\rm ren}$ approach to their classical value.
On the other hand, quantum fluctuation reduces $\Omega_{\rm ren}$ 
compared to its classical value, but enhances $\bar{\theta}_0$.
When the current is increased above a critical value, $j_c^{\eff}$,
no real physical solution to the self-consistent equation exists and
the parameters acquire imaginary part.
This means that the localized state in the well is no longer
metastable but becomes unstable due to zero point fluctuation and decays
directly from the well.
In the following, we will call this process as direct decay 
and its onset $j_c^{\eff}$ as effective critical current.

Now let us discuss the bias current dependences of the effective critical 
current,
$j_c^{\eff}$, and the decay rate, $\Gamma$, with $\alpha$ and $E_{J0}$ fixed.

To give a reference frame, we will compare 
the result with the approach where no retardation effect is taken into account.
One is the local and semi-classical approximation,
in which the kernel $K(\tau)$ is
treated as a $\delta$-function and the fluctuation of the phase is
neglected.(Denoted SC in the following.)
Second is the local approximation in which the quantum fluctuation are included
by self-consistent harmonic approximation as was done by Kleinert, but
the retardation effect of the kernel was neglected.(Denoted \lq 
without retardation'
in the following.)
Here the fluctuation around the self-consistent solution, which 
was taken into account in the Kleinert's original approach, is not considered.
(See section 4.3.)
The last is our approximation,
in which both the retardation effect of
the kernel and the quantum fluctuation are taken into account.
We fix the classical Josephson energy,
$E_J=\int_0^\beta K(\tau)\dtau$, for the above comparison.

It is to be noted that these three different theories treat 
the renormalization of the parameters due to the kernel, $K(\tau)$,
and quantum fluctuation very differently.
These renormalization are not considered in the first approach(SC), while
in the second treatment(\lq without retardation'), 
there is no renormalization to mass and
dissipation but the renormalization of frequency, 
since the kernel has no time dependence, and only
$\Omega_{\rm ren}$ and $\bar{\theta}_0$ are treated as variational parameters.

\begin{figure}
\epsfile{file=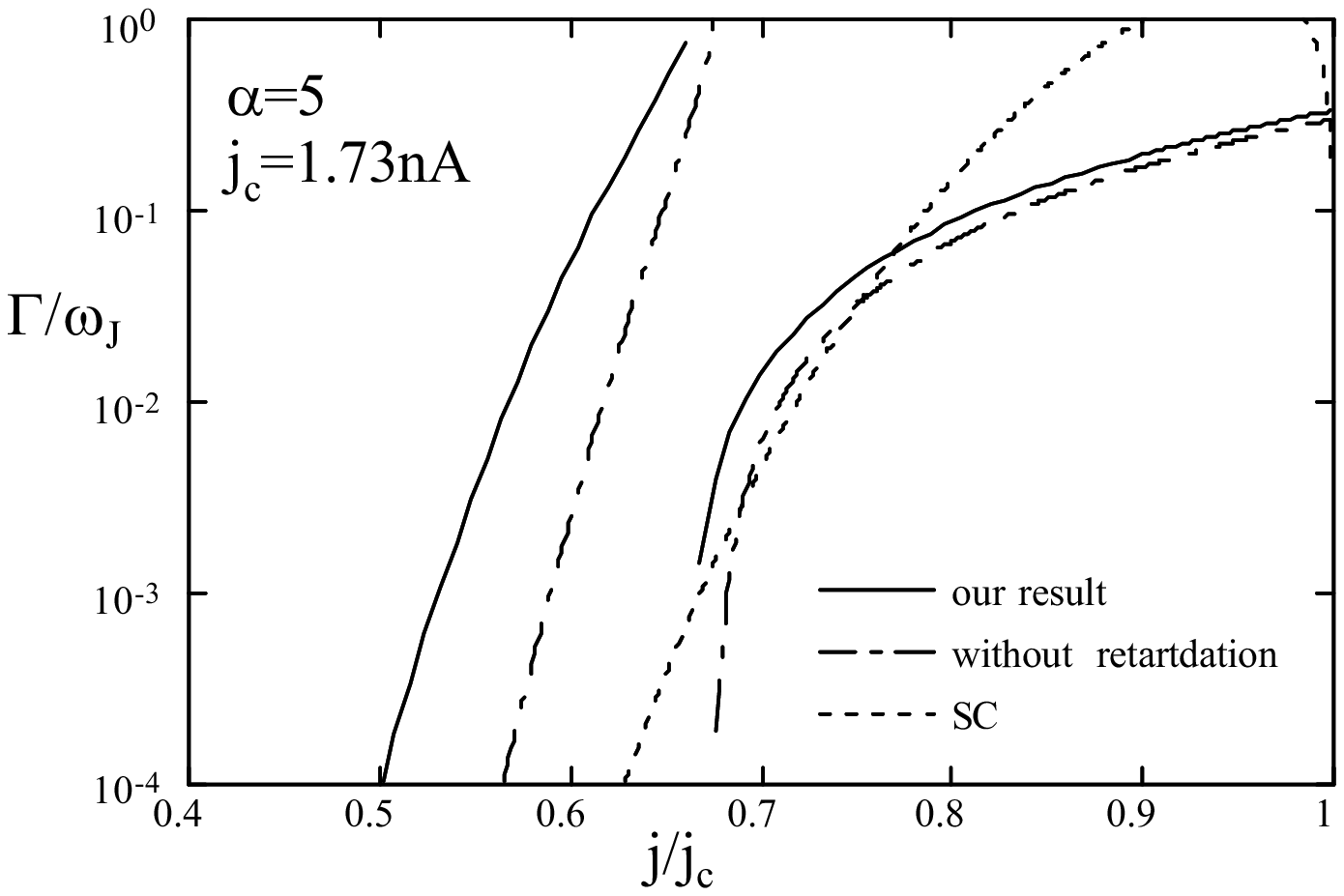,width=7cm}
\caption{Decay rate for point (a) in Fig.7 plotted as a function of $j/j_c$.}
\label{rate1}
\end{figure}

\begin{figure}
\epsfile{file=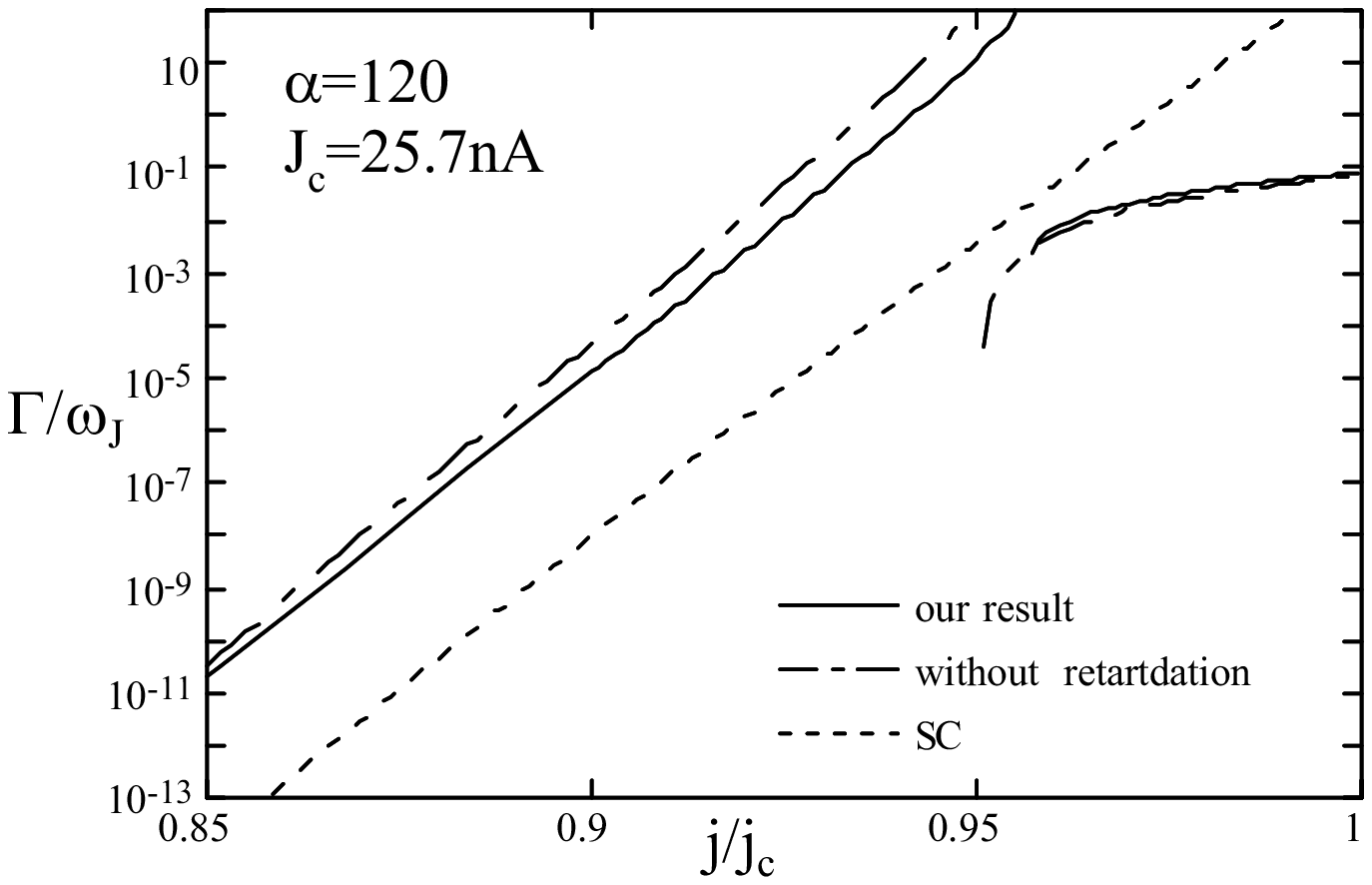,width=7cm}
\caption{Decay rate for point (d) in Fig.7 plotted as a function of $j$.}
\label{rate2}
\end{figure}

In Fig.~\ref{rate1} and Fig.~\ref{rate2}, $j$ dependence of
the decay rate $\Gamma$ is shown for the points (a) and (d), respectively.
(See also Fig.~\ref{phase})
Here, $\Gamma$ is plotted in unit of $\omega_J$. 
Since the parameter $\alpha$ is not large enough to use Korshnov's formula, 
we have used formula for cubic plus quadratic potential
for the tunneling rate in Fig.~\ref{rate1}.~\cite{CL2,FRH}
The solid line is our result, dotted and
the dot-dashed lines are by SC and \lq without retardation', respectively.

%#### Figure ####
%################

%#### Figure ####
%################

The overall behavior of the decay rate of our result is similar to that of
\lq without retardation'.
With the increase in $j$ toward $j_c^\eff$, decay rate increases. 
For current larger than $j_c^\eff$, there is no contribution from 
the bounce solution.
Instead the contribution from another process, {\it i.e.} the direct decay, 
sets in.
The direct decay rate is far smaller than the tunneling rate. 
This may partly be due to the overestimation of the tunneling rate in the 
localized regime as will be discussed later.

Next we discuss the difference between \lq without retardation'
 and our results in $j_c^{\eff}$ 
and $\Gamma$ in detail.
The $j$-dependence of decay rate for the parameters corresponding to the 
parameters of (d) in Fig.\ref{phase} is shown in
Fig.~\ref{rate2}.
Here we notice that $j_c^{\eff}$ is larger in our case than in 
\lq without retardation'.
This implies the smaller quantum fluctuation for non-local kernel for 
point (d). As a result $\Gamma$ is smaller in
our case than in \lq without retardation'. 
On the other hand, at point (a), we observe the opposite
behavior for $j_c^{\eff}$ and $\Gamma$.(See Fig.~\ref{rate1}.)
The different behavior seen in the quantum fluctuation of points
(a) and (d) can be considered as a result of competition between
the effect of $m_{\rm ren}$ and $\alpha_{\rm ren}$.
The renormalization effect makes $m_{\rm ren}$ larger and $\alpha_{\rm ren}$ smaller.
The former suppresses the quantum fluctuation and the latter enhances it.
For large $\alpha$, the former effect is dominant,
while for small $\alpha$ the latter effect becomes stronger and 
$j_c^\eff$ of our result is suppressed compared to that of 
\lq without retardation'.

\begin{figure}
\epsfile{file=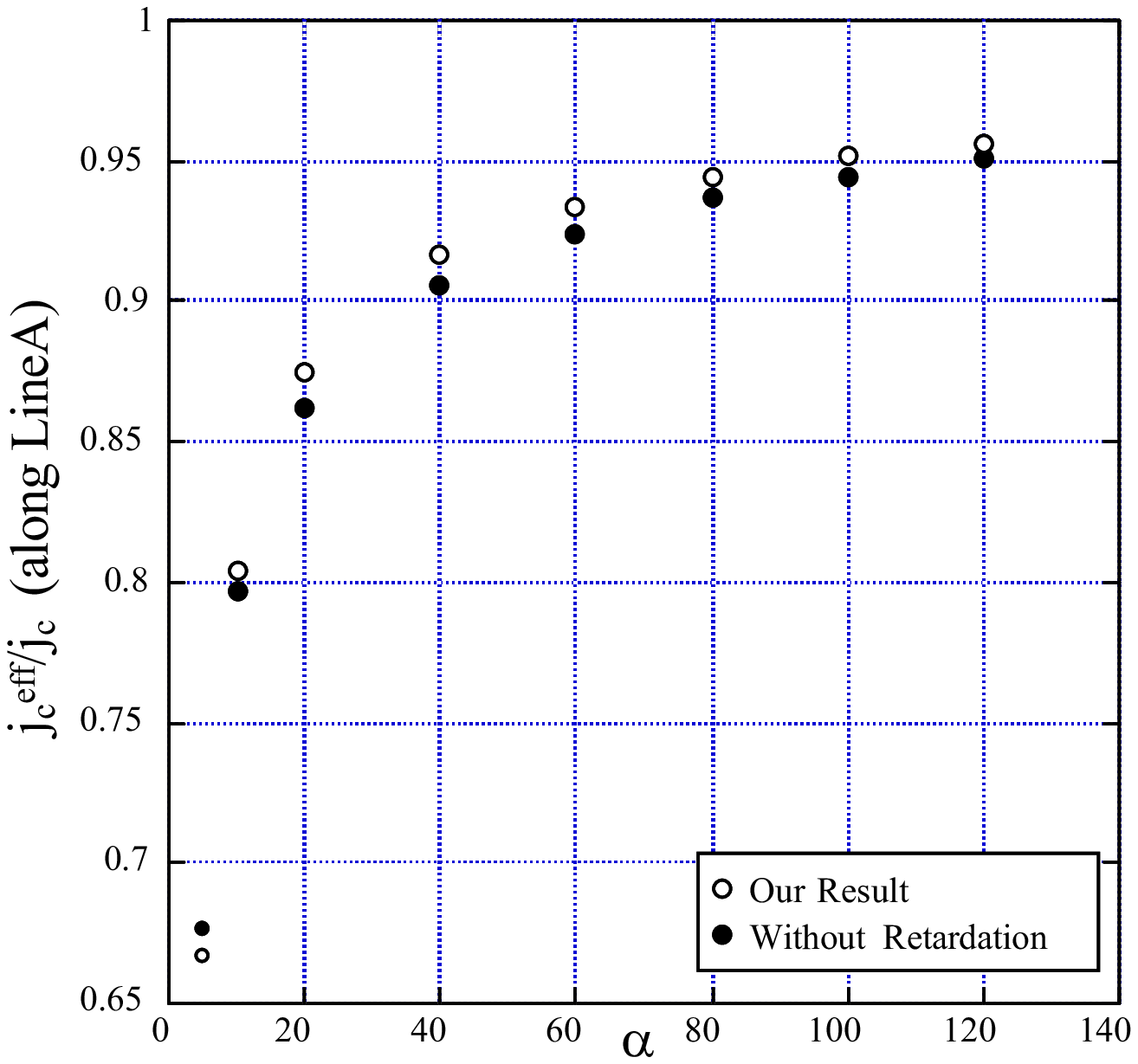,width=7cm}
\caption{The $\alpha$ dependence of the effective critical current along line A.
White circles denote our result and black ones the result of 
\lq without retardation'.}
\label{jcA}
\end{figure}

The $\alpha$ dependence of $j_c^{\eff}/j_c$ along line A is shown in 
Fig.~\ref{jcA}.
The white circles represent our result and the black ones for 
\lq without retardation'. 
We observe that the suppression of the critical current is not so large,
except in the neighborhood of localization-delocalization transition.
The dependence of $j_c^{\rm eff}$ on the parameter, $\alpha$,
is qualitatively the same as the classical one.
The difference in $j_c^\eff$ between our results and 
'without retardation' is also small.

%#### Figure ####
%################

\subsection{Discussion}
We conclude that the effect of the retardation of the kernel, $K(\tau)$, 
on the renormalization of $j_c^\eff$ is small compared to the case of local 
kernel $K(\tau)=E_J\delta(\tau)$.
The most significant effect, however, appears on the renormalization of 
mass and dissipation. 
The retardation effect leads to the $j$ dependent renormalization of 
mass and dissipation, which does not appear for the local kernel. 

Next we will discuss the limitation of our effective action.
As shown in the phase diagram in Fig.~\ref{phase}, we observe the
renormalized dissipation, $\alpha_{\rm ren}$, becomes negative in the shaded
region.
Since we have derived the effective action in the lowest order 
perturbation theory with respect to the matrix element of 
transfer integral of the electron at 
the SN boundary, the parameters $\alpha$ is given by the inverse resistance, 
$1/R_N$, of N region without coupling to the superconductor.
We expect the terms which compensates the 
negative contribution from the cosine term appears,
when the higher order terms in perturbation series are considered. 
Thus, we believe that the negative dissipation is an artifact 
of our derivation of effective action, in which only the lowest order 
terms are considered. 

Finally, we comment on the effect of fluctuation around the self-consistent 
solution. In our approach, tunneling rate, $\Gamma$, is evaluated with the 
renormalized potential obtained in the self-consistent harmonic approximation.
When the effect of fluctuation around the self-consistent solution is 
considered, the tunneling rate is expected to be reduced.
It is difficult, however, to treat this effect in our effective action, 
eq.(\ref{2.35}). Instead we here consider the simple case of cubic potential 
\lq without retardation', {\it i.e.}, 
\begin{equation}
V(x)={1\over2} m\omega_0^2x^2(1-{x\over x_0}).
\end{equation}
When $j\sim j_c$, this gives a good approximation to the cosine potential. 
The renormalized potential, ${\tilde V}(x)$, is given by,
\begin{equation}
{\tilde V}(x)={1\over2}m\Omega_{\rm ren}^2x^2-{1\over2}m{\omega_0^2\over x_0}x^3,
\end{equation}
where the effective attempt frequency, $\Omega_{\rm ren}$, is smaller than 
$\omega_0$. 
The imaginary part of free energy is calculated from the partition function,
\begin{eqnarray}
Z&=&\int\D x e^{-S}\nonumber\\
&=&\int\D x e^{-{\tilde S}} e^{-(S-{\tilde S})},
\end{eqnarray}
where $\tilde S$ is given by replacing $V(x)$ in $S$ with ${\tilde V}(x)$.
In the calculation of tunneling rate, we have evaluated ${\tilde S}$ by the 
bounce solution. Taking the fluctuation into account, 
we observe that additional factor, $e^{-\langle S-{\tilde S}\rangle_B}$, 
arises in the tunneling rate formula.
Here $\langle\cdots\rangle_B$ denotes the expectation value with respect to 
the fluctuation around the bounce path without the zero energy 
fluctuation. 
Since the expectation value $\langle y^2\rangle_B$ is positive, 
the expectation value of the factor,   
\begin{equation}
S-{\tilde S}
={1\over2}m(\omega_0^2-\Omega_{\rm ren}^2)\int_0^\beta\dtau (x_B+y)^2,
\end{equation}
is positive. Thus the tunneling rate is suppressed, when we consider the 
fluctuation around the self consistent solution. 
This situation is the same for $\phi^4$ potential\cite{Kleinert1}.
We also expect that this situation does not change 
when we consider the effective action for SNS junction  
and the tunneling rate is suppressed to be
closer to the direct decay rate.

\section{Summary}
In this paper, we have investigated the effect of low energy excitation 
due to conduction electron on the dynamics of the phase in SNS junction.
We derived the effective action for the phase in SNS junction from a 
microscopic Hamiltonian. 
The action is different from that of Josephson junction in two respects.
First, there is a term describing Ohmic dissipation expressed by 
the normal resistance of the junction, which is the resistance of the 
junction for large bias current limit. 
Second, the kernel, $K(\tau)$, which describes the Josephson coupling, 
has a long time tail. 
Both reflect the existence of low energy excitation in N region.
The adiabatic approximation cannot be justified {\it a priori} to this action,
and we focused on the effect of retardation of $K(\tau)$ on the dynamics of 
the phase. To clarify the effect, we have calculated the tunneling rate of 
the phase out of a minimum of potential with and without the retardation 
effect of kernel. 

The case where the N region is formed by the diffusive two dimensional 
electron gas was examined from the experimental interest.
The time dependence of $K(\tau)$ is characterized by the diffusion constant, 
$D$, which is proportional to the strength of the Ohmic dissipation $\alpha$.
Writing $K(\tau)$ as $K(\tau)\equiv E_{J0}{\hat k}(\tau)$, 
where $E_{J0}$ is a multiplication factor and ${\hat k}(\tau)$ is a function 
of $\alpha$, we have investigated dynamics of the phase by changing 
$E_{J0}$ and $\alpha$. 

The effect of the kernel was examined in the self-consistent harmonic 
approximation, and was expressed by renormalization of mass, $m_{\rm ren}$, 
dissipation, $\alpha_{\rm ren}$ and the attempt frequency, $\Omega_{\rm ren}$, 
of the phase variable. We have calculated the tunneling rate, $\Gamma$, of the 
phase by these renormalized parameters on the plane of $\alpha$ and 
$E_{\rm J0}$. These results are compared with the 
previous approaches; one is the semi-classical approximation(SC) in which 
neither of the retardation effect and quantum fluctuation is considered, and the other is the local approximation(without retardation), in which only the quantum fluctuation 
is considered. 

In large $\alpha$ limit, the motion of the phase is classical and our result 
approach to the semiclassical calculation. On the other hand for $\alpha<1$, 
the quantum fluctuation smears out the effect of cosine term. 
For the intermediate region, we found bias dependent renormalization of 
dissipation and mass.
The renormalization of dissipation makes the tunneling rate larger and 
that of mass makes it smaller compared to the unrenormalized case resulting in 
delicate competition of these two factors.
For a large value of dissipation, the latter effect dominates and the 
retardation effect makes the tunneling rate smaller.
For small value of the dissipation, {\it vice versa}.

\vspace{2em}
In conclusion, we have investigated the effect of retardation of the kernel 
$K(\tau)$ on the dynamics of the phase. 
We find the effect appears most significantly on the renormalization 
of mass and dissipation. 

Current dependent renormalization effect can be detected in the 
experiments of AC Josephson effect by measuring Q-value of plasma resonance 
directly.
It is possible to measure plasma frequency $\omega_p$ and the 
Q-value $\omega_p RC$ at the same time by applying small power rf voltage to 
the junction as in Dahm {\it et al.}~\cite{DDFLS}. 
Since $C$ and $R$ are proportional to $m_{\rm ren}$ and $1/\alpha$ 
respectively, it is expected to observe the reduction of $RC$ value as the 
increase of bias current in SNS junction.
As for the renormalization of $j_c^{\eff}$ and tunneling rate, 
$\Gamma$, however, the dependence to the time scale of the  $K(\tau)$ is 
found to be small.

\section*{Acknowledgements}
The authors acknowledge Dr. H. Takayanagi and Dr. H. Nakano 
for the informative discussions. 
They wish to express their gratitude to the Educational Computer 
Center of Tokyo University, where numerical part of this work was 
done.
This work is financially supported by Grant-in-Aid for Scientific 
Research on Priority Areas \lq\lq Quantum Coherent Electronics"
(06238103) from Ministry of Education, Science, Sports and Culture.
One of the authors(K. A.) appreciates the scholarship awarded by 
{\sc Texas Instruments Tsukuba Research and Development Center Ltd.}
\appendix
\section{Analytic Continuation of the Self Consistent Equation}
In this appendix, we will show the procedure to derive 
eqs.(\ref{sceq1}) from eq.(\ref{sceq}) by analytical continuation.

The self consistent equation in imaginary time is given as,
\begin{full} 
\begin{subeqnarray}
g_n^{-1}-g_{0n}^{-1}&=&\Sigma\pls {\rm i}\omega_n\prs\cos\bar{\theta}_0,\\
{j\over2e}&=&\Sigma\pls 0\prs \sin\bar{\theta}_0,\\
\Sigma\pls {\rm i}\omega_n\prs&\equiv&{1\over2}\int_0^\beta\dtau
K\pls\tau\prs\pls1+e^{-{\rm i}\omega_n\tau}\prs
e^{-{1\over4}\pl g\pls0\prs+g\pls\tau\prs\pr},
\label{A.1}
\end{subeqnarray}
\end{full}
where the function $g(\tau)$ and $K(\tau)$ are given by, 
\begin{full}
\begin{subeqnarray}
K\pls \tau\prs&=&
{\cal P}\int_{-\infty}^\infty {\dw\over2\pi {\rm i}}
{e^{-\omega|\tau|}\over 1-e^{-\beta\omega}}
\pll
K^{(R)}\pls \omega\prs
-K^{(A)}\pls \omega\prs
\prl,\nonumber\\
&=&
{\cal P}\int_{-\infty}^\infty {\dw\over\pi}
{e^{-\omega|\tau|}\over 1-{\rm e}^{-\beta\omega}}
\Im K^{(R)}\pls \omega\prs,\\
g\pls \tau\prs&=&{\cal P}\int_{-\infty}^\infty {\dw\over2\pi {\rm i}}
{e^{-\omega|\tau|}\over 1-{\rm e}^{-\beta\omega}}
\pll
g^{(R)}\pls \omega\prs
-g^{(A)}\pls \omega\prs
\prl.
\end{subeqnarray}
\end{full}
Our aim is analytic continuation of eq.(\ref{A.1}) from imaginary frequency 
to real frequency.
We first change the time variables as ${\rm i}z=\tau$.
With this, $\Sigma\pls {\rm i}\omega_n\prs$, becomes,
\begin{full}
\begin{eqnarray}
\Sigma\pls {\rm i}\omega_n\prs&=&{1\over2}\int_0^\beta\dtau K\pls \tau\prs 
\pls 1+e^{-{\rm i}\omega_n\tau}\prs
e^{-{1\over4}\pl g\pls 0\prs + g\pls \tau\prs\pr}\nonumber\\
&=&{1\over2}\int_0^{-{\rm i} \beta}
{\rm i}{\rm d} z K\pls {\rm i} z\prs \pls 1+e^{\omega_n z}\prs
e^{-{1\over4}\pl g\pls 0 \prs+g\pls {\rm i} z\prs\pr}.
\end{eqnarray}
\end{full}
Since there is no singularities in the region surrounded by 
$C$, $C_-$, $C_\infty$ and $C_\beta$(Fig.\ref{IC}),
we can change the integration contour as:$C\to C_-+C_\infty+C_\beta$ 
and obtain,
\begin{eqnarray}
\Sigma({\rm i}\omega_n)&=&{{\rm i}\over2}\int_{C_-+C_\infty+C_\beta}\dz
\pls 1+e^{\omega_n z}\prs e^{-{1\over4}g\pls0\prs}
\pll
K\pls {\rm i}z\prs e^{-{1\over4}g\pls {\rm i}z\prs}\prl,\nonumber\\
&=&{{\rm i}\over2}\int_{C_-+C_\beta}\dz
\pls 1+e^{\omega_n z}\prs e^{-{1\over4}g\pls0\prs}
\pll
K\pls {\rm i}z\prs e^{-{1\over4}g\pls {\rm i}z\prs}\prl.
\label{r.h.s.1}
\end{eqnarray}
Here the contribution from $C_\infty$ is zero.
The analytically continued form of the functions, $K({\rm i}z)$ and 
$g({\rm i}z)$, 
are given by, 
\begin{full}
\begin{subeqnarray}
K\pls {\rm i}z+\epsilon\prs&=&K_R\pls z\prs+{\rm i} K_I\pls z\prs,\\
K_R\pls z\prs&=&{\cal P}\int_{-\infty}^\infty {\dw\over\pi}
{\cos\omega z\over 1-e^{-\beta\omega}}
\Im K^{(R)}\pls \omega\prs,\\
K_I\pls z\prs&=&-{\cal P}\int_{-\infty}^\infty {\dw\over\pi}
{\sin\omega z\over 1-e^{-\beta\omega}}
\Im K^{(R)}\pls \omega\prs,\\
g\pls {\rm i}z+\epsilon\prs&=&g_R\pls z\prs+{\rm i} g_I\pls z\prs,\\
g_R\pls z\prs&\equiv&
{\cal P}\int_{-\infty}^\infty {\dw\over2\pi {\rm i}}
{\cos \omega z\over 1-e^{-\beta\omega}}
\pll
g^{(R)}\pls \omega\prs
-g^{(A)}\pls \omega\prs
\prl,\\
g_I\pls z\prs&\equiv&
-{\cal P}\int_{-\infty}^\infty {\dw\over2\pi {\rm i}}
{\sin \omega z\over 1-e^{-\beta\omega}}
\pll
g^{(R)}\pls \omega\prs
-g^{(A)}\pls \omega\prs
\prl.
\label{def}
\end{subeqnarray}
\end{full}
Note that when all the renormalized parameters are real, the definition 
$g_R$ and $g_I$ corresponds to the real part and imaginary part, respectively.
Otherwise, these functions have imaginary part.
Since $\Im K^{(R)}(\omega)$ is 
an odd function of $\omega$, we can show following relation easily,
\begin{subeqnarray}
K_R\pls z\prs&=&K_R\pls -z\prs,\\
K_I\pls z\prs&=-&K_I\pls -z\prs,\\
K\pls {\rm i}z+\beta-\epsilon\prs&=&K\pls {\rm i}z+\epsilon\prs,\\
K\pls {\rm i}z-\epsilon\prs&=&K_R\pls z\prs-{\rm i}K_I\pls z\prs.
\label{basic relation}
\end{subeqnarray}
The same relations as eq.(\ref{basic relation}) hold for $g({\rm i}z)$. 

With eqs.(\ref{def}) and (\ref{basic relation}), 
eq.(\ref{r.h.s.1}) becomes,
\begin{full}
\begin{equation}
\Sigma({\rm i}\omega_n)={1\over2} e^{-{1\over4}g\pls 0\prs }
\int_0^\infty\dz
\pls 1+e^{-\omega_n z}\prs
e^{-{1\over4}g_R\pls z\prs }
\pll
K_R\pls z\prs \sin{g_I\pls z\prs\over4}
-K_I\pls z\prs \cos{g_I\pls z\prs\over4}
\prl.
\end{equation}
\end{full}

By the analytical continuation, ${\rm i}\omega_n\to\omega+{\rm i}0^+$, 
we obtain,
\begin{equation}
g^{(R)}\pls \omega\prs=\pll g_0^{(R)-1}\pls \omega\prs
+\Sigma^{(R)}\pls\omega\prs\cos\bar{\theta}_0\prl^{-1}.
\label{self}
\end{equation}

At low frequency, we can expand eq.(\ref{self}), and obtain the 
self-consistent equations,
\begin{full}
\begin{subeqnarray}
m_{\rm ren}&=&m+{1\over4}e^{-{1\over4}g\pls0\prs}
\int_0^\infty\dz z^2 e^{-{1\over4}g_R\pls z\prs} 
\pll
K_R\pls z\prs \sin{g_I\pls z\prs\over4}
-K_I\pls z\prs \cos{g_I\pls z\prs\over4}
\prl\cos\bar{\theta}_0,\ \ \ \ \ \ \ \\
{\alpha_{\rm ren}\over2\pi}&=&{\alpha\over2\pi}
-{1\over2}e^{-{1\over4}g\pls0\prs}
\int_0^\infty\dz z e^{-{1\over4}g_R\pls z\prs}
\pll
K_R\pls z\prs \sin{g_I\pls z\prs\over4}
-K_I\pls z\prs \cos{g_I\pls z\prs\over4}
\prl\cos\bar{\theta}_0,\\
m_{\rm ren}\Omega_{\rm ren}^2&=&e^{-{1\over4}g\pls0\prs}
\int_0^\infty\dz e^{-{1\over4}g_R\pls z\prs}
\pll
K_R\pls z\prs \sin{g_I\pls z\prs\over4}
-K_I\pls z\prs \cos{g_I\pls z\prs\over4}
\prl\cos\bar{\theta}_0,\\
{j\over2e}&=&e^{-{1\over4}g\pls0\prs}
\int_0^\infty\dz e^{-{1\over4}g_R\pls z\prs}
\pll
K_R\pls z\prs \sin{g_I\pls z\prs\over4}
-K_I\pls z\prs \cos{g_I\pls z\prs\over4}
\prl
\sin\bar{\theta}_0.
\end{subeqnarray}
\end{full}

Hence eq.(\ref{sceq1}) is derived from eq.(\ref{sceq}).

%References

\end{document}